\documentclass[twocolumn]{aastex63}
\usepackage{xspace}
\usepackage{amsmath}
\usepackage[caption=false]{subfig}
\pdfoutput=1
\shorttitle{Classifying High-cadence Microlensing Light Curves I; Defining Features}
\shortauthors{Khakpash et al.}
\usepackage{natbib}
\usepackage{hyperref}

\begin{document}

\title{Classifying High-cadence Microlensing Light Curves I: Defining Features}

\author[0000-0002-1910-7065]{Somayeh Khakpash}
\affiliation{Department of Physics, Lehigh University, 16 Memorial Drive East, Bethlehem, PA 18015, USA}

\author[0000-0002-3827-8417]{Joshua Pepper}
\affiliation{Department of Physics, Lehigh University, 16 Memorial Drive East, Bethlehem, PA 18015, USA}

\author[0000-0001-7506-5640]{Matthew Penny}
\affiliation{Department of Physics \& Astronomy, 261-B Nicholson Hall, Tower Dr., Baton Rouge, LA 70803, USA}

\author[0000-0003-0395-9869]{B. Scott Gaudi}
\affiliation{Department of Astronomy, The Ohio State University, 140 W. 18th Ave., Columbus, OH 43210}

\author{R. A. Street}
\affiliation{LCOGT, 6740 Cortona Dr, Goleta, CA 93117, USA}

\date{\today}

\begin{abstract}
\noindent Microlensing is a powerful tool for discovering cold exoplanets, and the The Roman Space Telescope microlensing survey will discover over 1000 such planets. Rapid, automated classification of Roman's microlensing events can be used to prioritize follow-up observations of the most interesting events. Machine learning is now often used for classification problems in astronomy, but the success of such algorithms can rely on the definition of appropriate features that capture essential elements of the observations that can map to parameters of interest. In this paper, we introduce tools that we have developed to capture features in simulated Roman light curves of different types of microlensing events, and evaluate their effectiveness in classifying microlensing light curves. These features are quantified as parameters that can be used to decide the likelihood that a given light curve is due to a specific type of microlensing event. This method leaves us with a list of parameters that describe features like the smoothness of the peak, symmetry, the number of peaks, and width and height of small deviations from the main peak. This will allow us to quickly analyze a set of microlensing light curves and later use the resulting parameters as input to machine learning algorithms to classify the events.
\end{abstract}

\section{Introduction} \label{sec:intro}

Microlensing is a phenomenon that happens when light emitted from a distant object (the source) is lensed by a closer, massive object (the lens), and as a result, multiple images of the source are formed. These images are typically not resolved because their angular separation is much smaller than the angular resolution of both ground- and space-based telescopes, and consequently, we observe a brightening of the source.

Microlensing is a powerful tool for detecting small and dim objects that are otherwise very hard to detect by their emitted light, and in particular, it is so far the only method capable of investigating planetary systems with terrestrial-mass planets orbiting beyond the snow line \citep{gaudi2010exoplanetary}.  For several decades, ground-based surveys have searched the Galactic Bulge for microlensing events. The Optical Gravitational Lensing Experiment (OGLE) \citep{udalski2015ogle}, the Microlensing Observations in Astrophysics (MOA) \citep{bond2001real} and the Korea Microlensing Telescope Network (KMTNet) \citep{kim2016kmtnet} surveys detect thousands of microlensing events. These observations have led to the successful discovery of over 100 exoplanets\footnote{https://exoplanetarchive.ipac.caltech.edu} and some compact substellar objects \citep{griest1995macho,bennett2002gravitational,mao2002optical}.

Identifying microlensing events from the huge numbers of light curves obtained in these surveys is a challenge. It is useful to be able to distinguish microlensing events either in real time during survey operations as the events start, or after an observing campaign has finished and the data of being fully analyzed.  Microlensing surveys have generally relied on circulating real-time alerts of potential microlensing events when there is an increase in the brightness of an observed source.  This method can lead to alerts for any sudden rise in a light curve, which can include variability other than microlensing, such as cataclysmic variables (CVs). There has been recent work to increase the accuracy of these alerts, limiting them to genuine microlensing events, using machine learning (ML) algorithms that can distinguish microlensing light curves from other variability in real time \citep{kessler2019models,godines2019machine}. Real-time classification can increase the accuracy of alerts.  It can also identify high-value events for which only partial coverage will be obtained by the survey, and allow observers to schedule supplemental follow-up observations to increase the coverage of the event.

There is also a need for classification of microlensing survey data after the conclusion of observations.  At that point, it is necessary to first distinguish microlensing events from other types of variability, and also to separate different types of microlensing events, such as those that include planetary signals. Surveys like the KMTNet survey have employed an automated approach to detect microlensing-like variability in complete light curves by fitting simple functions \citep{kim2018korea}. \citet{belokurov2003light} also advocates the use of neural networks to distinguish well-sampled microlensing light curves from other types of variability. In this paper, we wish to explore the efficiency and efficacy of after-the-fact classification of microlensing signals.

ML has now become a popular method for classifying astronomical time series. Some ML classifiers like the Random Forest \citep{liaw2002class} and k-mean \citep{lloyd1982least} classifiers take light curve features as input and try to find a connection between those features and the classification labels. In this scenario, ``features" are quantitative statistical or morphological measurements of the time series. Some other methods like neural networks use the light curves themselves as input, and then find common patterns or higher-order correlated properties to classify them. As a specific example,  Random Forest is a widely-used algorithm that classifies time series by making decisions based on features in the time series \citep{bluck2020galactic,pawlak2019machine}. In order to use this algorithm efficiently, observable features in the light curves that are most closely related to the canonical model parameters must first be identified.

Here, we focus on analyzing complete high-cadence microlensing light curves. These are simulated light curves for the Roman Galactic Bulge Exoplanet Survey. The Nancy Grace Roman Space Telescope (Roman), formally known as the Wide Field Infrared Survey Telescope (WFIRST) is a NASA future space mission that is expected to be launched by mid 2020's. One of its primary goals is to detect exoplanets using the microlensing method \citep{spergel2015wide}. It is estimated that Roman will find about $54,000$ microlensing events and will detect $1400$ planets \citep{penny2019predictions}. The $0.01$ AU distance of Roman from the Earth enables measuring parallaxes for free-floating planets and Earth-mass bound planets to help constrain their masses \citep{zhu2016augmenting,street2018unique}. To assess the value of ML techniques to these surveys we will apply ML features and classifiers to simulated Roman microlensing data produced for the Roman microlensing data challenge\footnote{https://microlensing-source.org/data-challenge/} (Street et al. in prep.).

Binary-lens light curves are often difficult to model because of the complicated features in their light curves and the large parameter space that needs to be fully searched (for a more thorough review refer to the introduction of \citet{penny2014speeding} and \citet{khakpash2019fast}).  Those challenges highlight the importance of developing fast automated algorithms to quickly analyze light curves and estimate the microlensing model parameters. Our primary goal in this work is to identify a list of features specifically defined for microlensing light curves that can be generated by fast and efficient algorithms, and show that these features can help either in differentiating microlensing light curves among other types of variability or in classifying microlensing light curves into different types. This would enable fast detection of planetary system lenses and other interesting lensing cases in the released datasets of large surveys like Roman.

We present a collection of algorithms including various functional fits that are applied to the light curves, and from these fits, we extract parameters that quantify features of the light curve like smoothness of the peak, symmetry, number of peaks, similarity to microlensing single-lens curves, number of deviations from the peak, and width and height of the deviations from the main peak. We then show how effective each of these functions are in distinguishing between different types of microlensing events like identifying single-lens versus multiple-lens systems, or planetary system lenses versus stellar binary lens systems, and in some cases in detecting microlensing events among other types of variability. A similar work was first done by \citet{mao1994interpretation} where they used features such as an estimate of the asymmetry about the peak to detect binary-lens signatures in the light curves. In Section \ref{sec:types}, we introduce the different models of microlensing events that need to be parameterized for the classification, and in Section \ref{sec:scope}, we discuss the properties of our test dataset. In Section \ref{sec:define}, we introduce our algorithm package including the different functional fits and their respective output parameters. We also evaluate the effectiveness of each function in capturing the specific set of features in the light curve. In Section \ref{sec:ML_test}, we show preliminary tests of using these features as input to machine learning classifiers. Finally, in Section \ref{sec:conclusion}, we discuss our results and applicability of our method to other datasets.

\section{Microlensing Models}\label{sec:types}

\begin{figure*}[!tbp]
    \centering
    \includegraphics[scale=0.27]{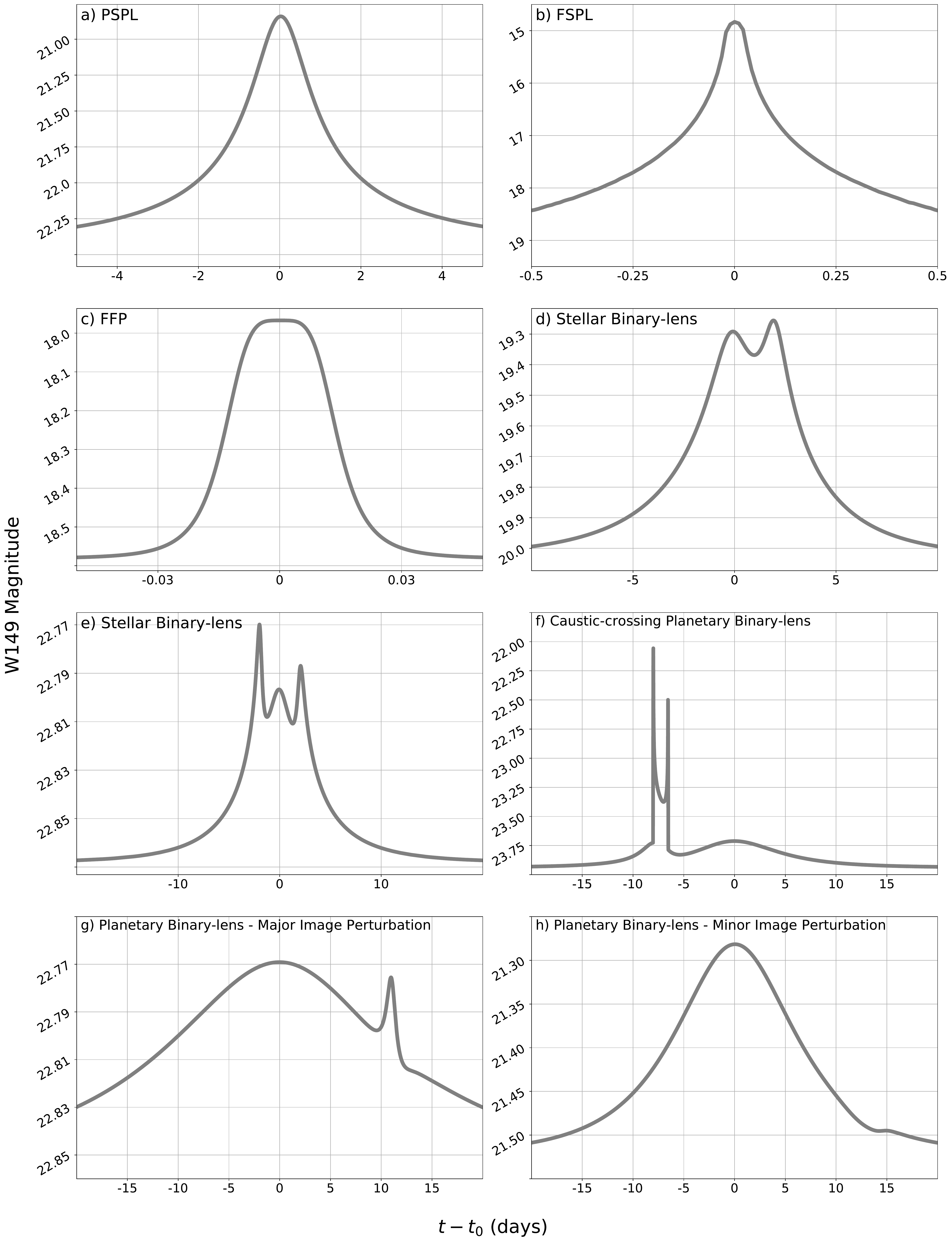}
    \caption{Light curves containing different types of major features in microlensing light curves. The top row includes light curves created by single lenses. Panel (a) is a single stellar lens with no significant finite source effect. Panel (b) shows the effects of the finite source effect on a single-lens stellar event. Panel (c) shows an event due to a free-floating planet. Panels (d) and (e) show two examples of events caused by binary star lens systems. Panels (g) and (h) are due to planetary systems and have no caustic crossings, whereas Panel (f) is a planetary event that contains a caustic-crossing event.}
    \label{fig:lc_features}
\end{figure*}

Microlensing occurs when the light coming from a distant source is lensed by a closer object along the line of sight. As a result, the source appears brighter as the angular separation of the two objects decreases. This phenomenon then results in a peak in the light curve of the source star at the time of closest angular approach. A simple single-lens microlensing light curve has a single symmetric peak as shown in panel (a) of Figure \ref{fig:lc_features}. The different panels in Figure \ref{fig:lc_features}, represent different shapes of microlensing events that might be present in a large dataset of a galactic bulge survey. Note that it is practically impossible to include all possible types of light curve morphologies and this plot is only a small subset of all possibilities.

The light curves in Figure \ref{fig:lc_features} have characteristics representative of the types of features we are seeking to automatically detect in this work. When there are multiple lenses or sources, and when there are second-order effects like finite source effect and parallax, the shape of the single-lens peak will deviate from a symmetric peak as in panel (a). In this section, we will discuss the different  physical phenomena that can cause these deviations.

Panels (b) and (c) in Figure \ref{fig:lc_features} show two single-lens models affected by the finite source effect \citep{witt1994can, nemiroff1994finite, gould1996finite}. Panel (c) contains an event caused by a free-floating planet and therefore the light curve is more affected by the finite source effect. Panels (d) and (e) include examples light curves of binary-lens events caused by binary stars, and panel (f) has an example of a planetary binary-lens microlensing event with a caustic crossing. Panels (g) and (h) have examples of planetary binary-lens events with no caustic crossings. The example in panel (g) is an example of a major image perturbation and the event in panel (h) is a minor image perturbation (for a discussion of major/minor images refer to Section \ref{subsub:Planetary_binary}). In the following subsections, we introduce different physical models that give rise to the light curve features seen in Figure \ref{fig:lc_features}. 

\subsection{Point-source Point-lens Microlensing Light Curves}\label{subsec:PSPL}

\begin{figure}[!ht]
    \centerline{\includegraphics[width=8cm]{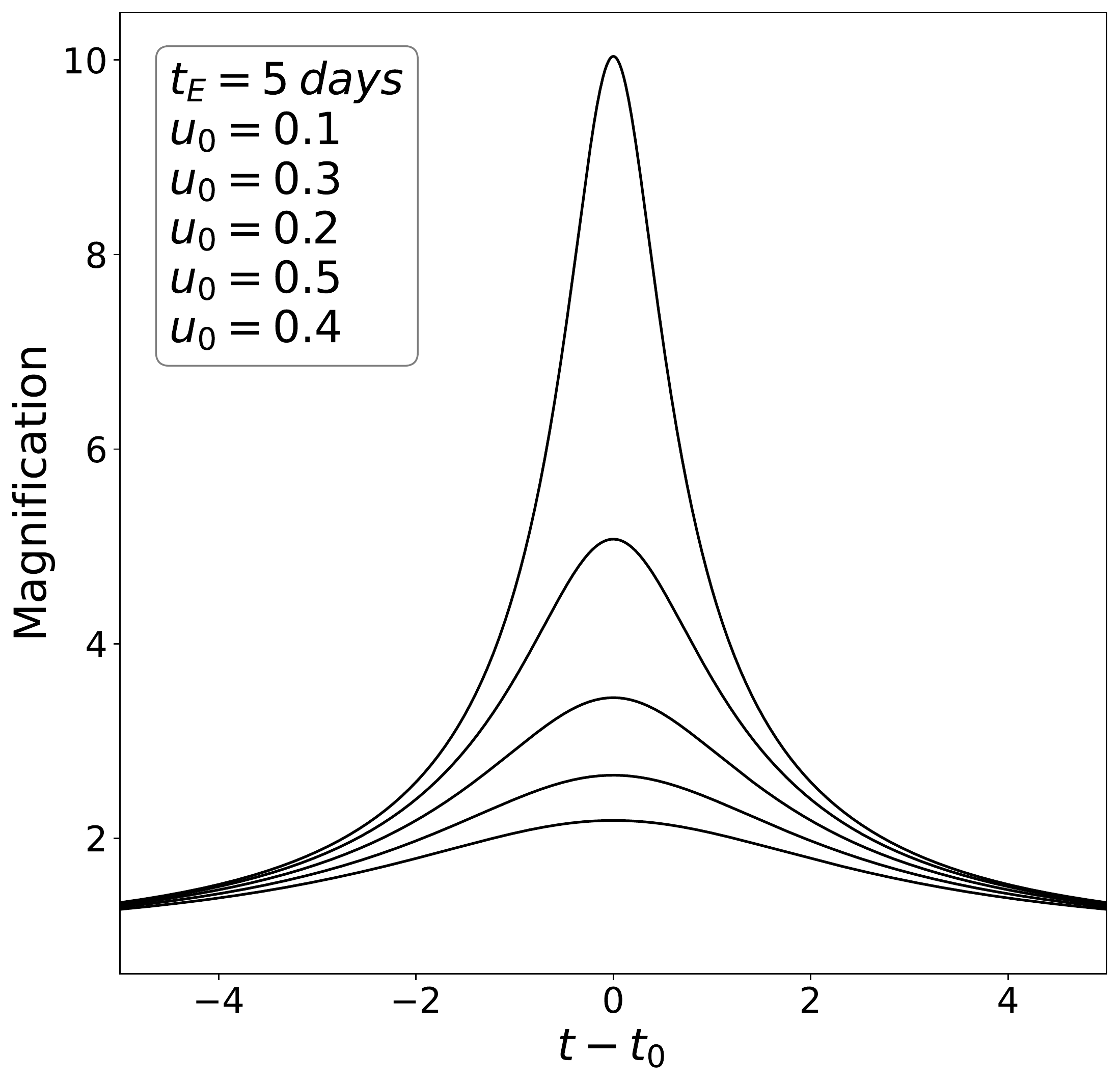}}
    \caption{Five PSPL models with different values of $u_0$. The figure shows how the shape of the curve changes for different values of the impact parameter.}
    \label{fig:PSPL_u0}
\end{figure}

The Point-source Point-lens (PSPL) model of a lensing event when there is a single lens and a single source, and the source and the lens can both be approximated as point-like objects with zero angular size. 
When there is no blending of the source with the neighboring stars or light from the lens or companions to the lens or source, this model can be described by a simple Paczy\'nski curve as in Equation \ref{eqn:A_PSPL} and \ref{eqn:u}. $A$ is the magnification of the source, $t_0$ is the time of the maximum magnification, $u_0$ is the impact parameter, and $t_E$ is the Einstein crossing time. Figure \ref{fig:PSPL_u0} shows five different PSPL models with the same Einstein timescale, and different values of $u_0$.  In the presence of blending, another parameter is added to Equation \ref{eqn:A_PSPL} and will yield Equation \ref{eqn:PSPL_A_fs} where $F(t)$ is the differential flux and $f_s$ is the blending parameter and determines what fraction of the total flux in the aperture is due only to the source. When there are no higher order effects, these curves remain symmetric, and the sharpness of the peak in high-magnification events remains intact. Detecting any deviation from this model can help identify the presence of higher order effects.

\begin{equation}\label{eqn:A_PSPL}
    A(t) = \frac{u(t)^2 + 2}{u(t)\sqrt{u^2 + 4} }
\end{equation}
\begin{equation}\label{eqn:u}
    u(t) = \sqrt{{u_0}^2 + {\left(\frac{t-t_0}{t_E}\right)}^2}
\end{equation}
\begin{equation}\label{eqn:PSPL_A_fs}
    F(t) = f_s \times A(t)+(1-f_s) 
\end{equation}

\subsection{Finite Source Effects}\label{subsec:Finite}

When the source is not point-like, and has a finite size, the magnification function will be different, especially when the impact parameter is comparable to the source size \citep{witt1994can}. In Figure \ref{fig:lc_features}, an example of the changes due to finite source effects can be seen in panel (b). For small impact parameters, this effect becomes stronger, and it smooths out the peak of the event. This effect can be used to estimate $\theta_E$, the angular size of the Einstein ring and therefore provide constraints on the lens mass \citep{yoo2004ogle,jiang2004ogle}.

\subsection{Free Floating Planets}\label{subsec:free}

Isolated, short duration microlensing events can be due to free floating planets (FFPs) or planets in very wide orbits. Compred to most other exoplanet detection methods, microlensing has the capability to detect these planets since the method only depends on the mass of the lens and not its luminosity. Microlensing events caused by FFPs have short timescales of $t_E < 2$ days because the size of the Einstein ring depends on the lens mass, and they can be detected in high-cadence observations \citep{sumi2011unbound,mroz2017no,mroz2018neptune,mroz2020free}. Roman's unique combination of high cadence and small photometric noise makes it particularly favorable for detecting the free-floating planets \citep{johnson2020predictions}. 

Microlensing due to FFPs is very likely to be affected by finite source effects. The finite source effect \citep{witt1994can} parameter, $\rho$, is defined as ${\theta_{\ast}}/{\theta_E}$ where $\theta_{\ast}$ is the apparent size of the source, and $\theta_E$ is the angular size of the Einstein ring. $\theta_E$ depends on the lens mass, and it becomes smaller for single planetary lenses, and therefore, $\rho$ becomes larger, and the finite source effect is stronger. As a result, the top of the short-timescale peak becomes flattened, and the overall shape of the event begins to resemble a tophat function. An example of such an event can be seen in panel (c) of Figure \ref{fig:lc_features}.

\subsection{Binary-lens Microlensing Events}\label{subsec:Binary}

When there is more than one lens, the range of possible light curve morphologies increases dramatically. In binary-lens systems, various shapes of the light curves depend on parameters such as $q$, the mass ratio, $s$, the projected separation, and $\alpha$, the angle the path of the source makes with the line connecting the two lenses. Since $\alpha$ can be any value between zero and $360^{\circ}$, this quantity along with $q$, $s$, and $t_E$ allow the time, height, width, and number of the features in the light curve vary significantly. Based on OGLE \citep{udalski2004optical} and MOA \citep{bond2001real} observations, around 10\% of microlensing light curves show lens binarity signatures \citep{penny2011detectability}. Fully modeling each of these light curves is challenging because there are multiple minima in the $\chi^2$ surface as a function of their parameters, and therefore a wide range of parameter space must be searched. Furthermore, there is no direct mapping between the observable features and the canonical parameters ($q,s,\alpha,t_0,t_E,u_0$).

\subsubsection{Stellar Binary-lens Events}\label{subsub:Stellar_binary}

Stellar binary lenses are likely as common or more common than planetary system lenses, but the probability of seeing deviations from a PSPL curve is much higher in nearly equal mass ($q \gtrsim 0.1$) binary lens light curves than light curves due to planetary mass ratio ($q \lesssim 0.01$) binaries because the size of the caustics (the loci of infinite magnification for a point source) increase with increasing $q$. Therefore, they are typically easy to distinguish from planetary system lenses since the deviations from the standard PSPL form last much longer, and in particular can be comparable to $t_E$. They usually have larger perturbations in their light curves, but can be still misinterpreted as planetary binary-lens events \citep{han2016new}. Two examples of such light curve can be seen in panels (d) and (e) of Figure \ref{fig:lc_features}.

\subsubsection{Planetary Binary-lens Events}\label{subsub:Planetary_binary}

A lens system containing a star and a single planet is a binary-lens system with a very small mass ratio ($q$). Note that there is no strict delineation (in terms of $q$) between a stellar binary-lens event and a planetary binary-lens event, and this is one of the reasons that make this a challenging classification problem.  In these cases, the light curve is dominated by the host star, such that there is a main peak which can be described by the PSPL model with one or more small deviations caused by the planet \citep{gaudi2012microlensing}. As in a single-lens microlensing event, two images of the source will be formed as it passes close by the lens star; an image outside of the Einstein ring of the lens star called the major image, and an image inside the Einstein ring of the lens star called the minor image. Depending on whether or not the projected separation of the planet and the lens star ($s$) is larger or smaller than the Einstein ring, the planet will have to perturb one of these images to leave a signature on the star light curve, and the perturbation will have different characteristics. The effect of the planet on the light curve can also be explained by the positions of caustics. In real cases, since the source star is not a point source, we observe sharp high-magnification peaks when there are caustic crossings. The numbers and sizes of these caustics depend on the mass ratio ($q$) and the projected separation ($s$). The structure of the caustics leads to the planetary features in the light curves of these events \citep{gaudi2012microlensing}. 

Caustics have three main topologies. Depending on the values of the projected separation and the mass ratio, there is either one (intermediate topology), two (wide topology), or three (close topology) caustics (for a thorough description of planetary caustics please refer to \citet{gaudi2012microlensing}). These caustics are also classified into two classes, called the central caustic and the planetary caustic, and depending on which one is closer to the path of the source, the shapes of the light curve are different.

If the source crosses a caustic curve, two sharp deviations will be seen in the light curve assuming sufficient sampling corresponding to the entrance and exit of the source through the caustic. The shape of these deviations depends on the path of the source which is determined by the source path angle ($\alpha$). Panel (f) of Figure \ref{fig:lc_features} includes an example of a caustic-crossing planetary microlensing event. In this work, we analyze these perturbations based on whether or not the source has crossed the caustics, and we investigate the features created in each of these cases.  
In cases with no caustic crossing, the source only passes close to the caustics, and that results in features like short bumps or troughs, or both, in the light curve. These deviations are typically small in magnitude and short in duration which makes them sometimes very difficult to detect, especially in low-cadence surveys, and they can be created by either the central or the planetary caustics. In panels (g) and (h) of Figure \ref{fig:lc_features}, we show examples of non-caustic-crossing events caused by the planetary caustic. These perturbations happen when the source passes closely by the planetary caustic, and finding the time of the perturbation can help determine the location of the caustic, and therefore estimate the projected separation of the planetary system. The event in panel (g) is a perturbation in the major image and the event in panel (h) is a perturbation in the minor image.

In cases with caustic crossing, we need to identify the times of the two sharp deviations to find an estimate of the time the source spends inside the caustic. This allows us to estimate the size and location of the caustics, and therefore find an estimation of the mass ratio and the projected separation of the planetary system \citep[e.g.,][]{poleski2014super}.

\section{Scope of the Analysis}\label{sec:scope}

\begin{figure*}[t]
    \centerline{\includegraphics[scale= 0.4]{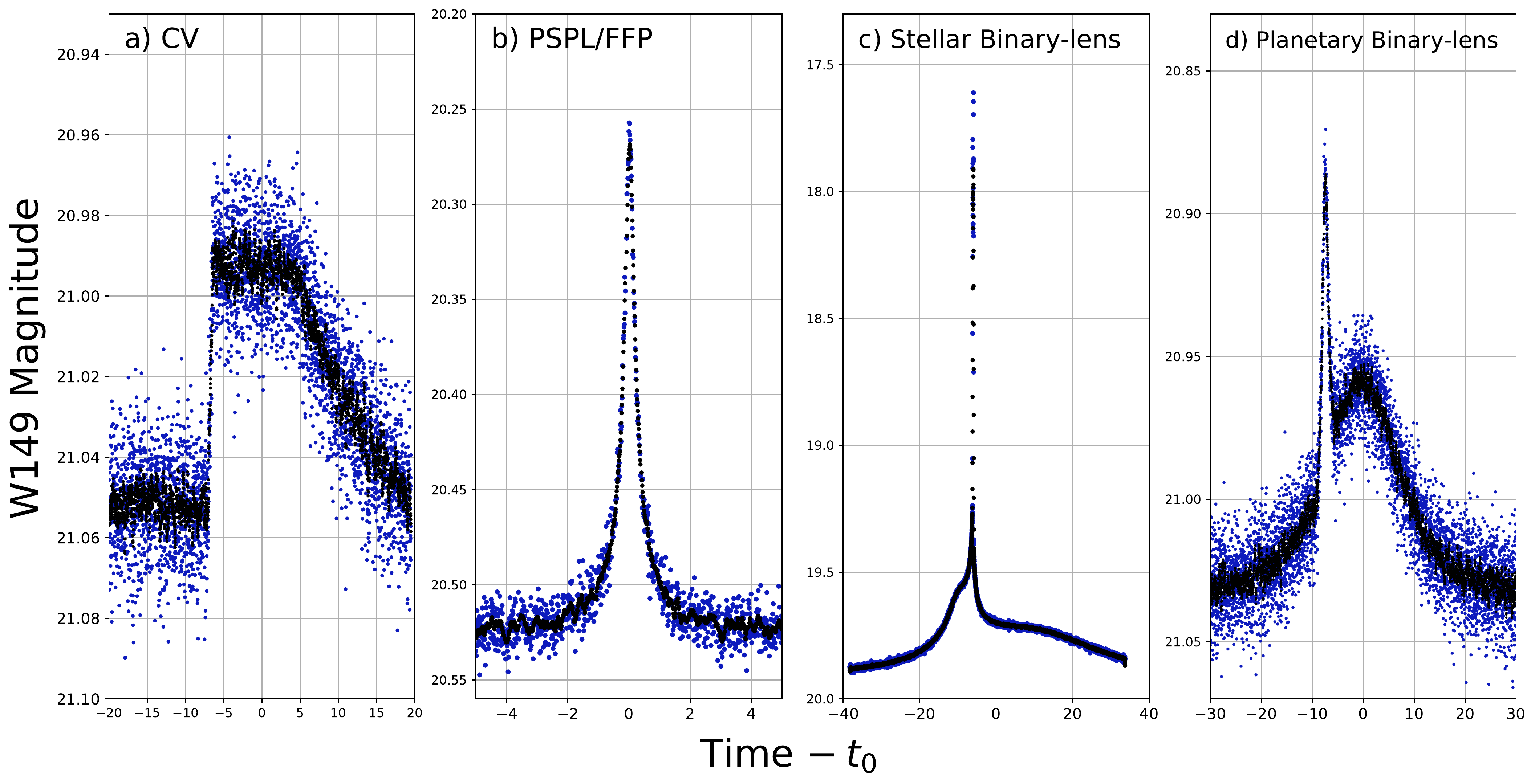}}
    \caption{Four classes of light curves in the dataset used for this analysis. From left: is an example of the light curve of a cataclysmic variable. The second panel shows a single-lens model which contains a single symmetric peak. The third panel represents lens systems containing a stellar binary, and the last panel is an example of an event with a planetary system as the lens. The blue data points are from the original simulated light curves, and the black data points are smoothed light curves using the low-pass filter.}
    \label{fig:lc_classes}
\end{figure*}

As mentioned in Section \ref{sec:intro}, classifying astronomical time series using ML algorithms relies on the data sets used for training and testing. Features computed for a data set cannot always be used to analyze another data set; therefore, the connection between the dataset and the defined features should be discussed. In this paper, we focus on computing features for simulated Roman light curves. These light curves are simulated based on the Roman Cycle 7 mission design and have six 72-days seasons with 15 minute cadence. These light curves are similar to the light curves generated by \citet{penny2013exels} and \citet{penny2019predictions}, and the full details of the simulated data can be found in these papers. This data set was designed in particular to be used for the Roman microlensing data challenge\footnote{https://microlensing-source.org/data-challenge/} and has four classes of variability types, consisting of different types of microlensing light curves along with a particular type of stellar variability that can be misinterpreted as microlensing. These classes include cataclysmic variables, single-lens systems, stellar binary lenses and planetary system lenses. An example of each class of light curves used in this analysis can be seen in Figure \ref{fig:lc_classes}. 

The dataset contains $4181$ light curves, including $429$ cataclysmic variables, $1626$ single-lens systems, $1386$ stellar binary lenses, and $688$ planetary system lenses. The CV class contains CV variabilities with either zero, one or multiple instances of outbursts in the baseline. The single-lens systems include both stars and planets as lenses, therefore their shape varies in terms of the duration, amplitude, and morphology of the peak. As mentioned before, the stellar binaries and the planetary systems can result in very similar light curves, and the transition between these two classes is smooth in terms of the observables, and distinguishing between them is one of our main concerns. 

The lightcurves selected for this project have a wide range of signal to noise ratio. To enter our sample, events had to pass a $\Delta{\chi}^2 >500$ cut of the true model to a flat line, and binary and planetary events had to pass a ${\Delta \chi}^2 > 160$ cut of the true model relative to a single lens lightcurve model (see \citet{penny2019predictions} for full details). While these ${\Delta \chi}^2$ values suggest quite high formal detection significance, the cuts can be passed by lightcurves with signal to noise per data point less than unity, e.g., a $t_{\rm E}=25$ day single lens event will have 4800 data points within one $t_{\rm E}$ of the event peak. Similarly, a Jupiter-mass planet with a planetary deviation lasting ${\sim}1$ day (i.e., ${\sim}96$ data points) can pass the detection cut with an average contribution of  ${\Delta \chi}^2$ per data point of less than 2. This said, our selection cuts do not allow us to test our method's sensitivity to the lowest signal to noise ratio events that could potentially be formally detected.

\section{Feature Detection and Parametrization}
\label{sec:define}

In Section \ref{sec:types}, we described major types of microlensing variability with different morphologies and amplitudes. In this section, we introduce a package of statistical tests and model fits we have assembled into an algorithmic approach to detect the features of a set of light curves. These features individually can then be used to distinguish microlensing events from other variability, and to classify the different types of microlensing systems. Ultimately, in a subsequent paper, we will use the combined power of all of the features as input to a ML classifier. The package includes the following tools: 

\begin{itemize}
    \item Peak Finder
    \item Gould Two-parameter Point-source Point-lens Fit (G-PSPL)
    \item Symmetry Check
    \item Trapezoidal Function Fit
    \item Cauchy Distribution Fit
    \item Planetary Parameters Finder
    \item Chebyshev Polynomials Fit
\end{itemize}

In the sections below we introduce each of these tools, and explain how they are applied to the simulated data set and evaluated. Not all tools are used independently, and some are used in combination with other tools.

Before applying some of these tools, we employ a low-pass filter to the light curve which allows low frequency signals to pass and therefore reduces the high-frequency noise in the data. This low-pass filter has a smoothing window of 10 subsequent data points. The width of the smoothing window was optimized for the simulated Roman data set analyzed in this work, but should be separately optimized for each data set so as to remove non-astrophysical noise as much as possible. 

\subsection{Peak Finder} \label{subsec:PeakFinder}

The primary idea behind this tool is to identify individual peaks in a light curve, which are often characteristic of microlensing events. After employing the low-pass filter, we bin the light curve in time, and then, in each bin, we count the number of data points with positive deviations larger than some number of standard deviations in that bin, which we define as the peak threshold. If a bin has more than one data point beyond that threshold, that bin is defined as a peak.

When using this algorithm, two parameters are important in determining the success rate; the bin size and the peak threshold. Depending on the particular implementation, we can use this method to search for a single large peak, such as a single-lens microlensing event. Or after identifying a single-lens event, we can subtract a model and then search for additional peaks in the light curve residuals that would indicate a binary lens event.  We can also search simultaneously for multiple large peaks at once.

To examine the performance of this method in a particular case, we consider the case of searching for a single peak within a light curve. We test different values of the bin size and threshold parameters to determine what combination is more successful at detecting peaks. We define a successful detection as one where exactly one bin identifies a peak, and the light curve is known to be due to a microlensing event.  We define a failed detection as one where a light curve labeled as containing a microlensing event did not have any peaks identified, or multiple peaks identified. With this range of bin sizes, this algorithm will not detect narrow deviations in microlensing light curves, and will only look for the main event. Note that the algorithm is looking for single-peaked variations, so it might also find single-peaked transients. We expect that next stages of analysis will distinguish between these cases. This method has a low false negative rate. For example, it is possible that it would fail at detecting a microlensing light curve at low thresholds. This happens when the event has a low magnification and it might be considered similar to the noise in the light curve.

\begin{figure}
    \centerline{\includegraphics[width=1\linewidth, clip]{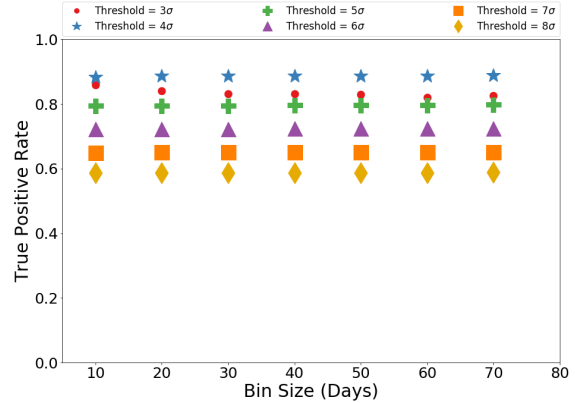}}
    \caption{Results of the Peak Finder algorithm is applied to all of the light curves in the dataset for different values of bin size and threshold. The True Positive Rate (TPR), the fraction of microlensing events that are labeled as microlensing is plotted versus bin size, with various thresholds denoted by symbol and color. The value of TPR is roughly independent of the bin sizes, and that thresholds of $3\sigma$ to $5\sigma$ yield the highest TPR. }
    \label{fig:TPR}
\end{figure}

\begin{figure}
    \centerline{\includegraphics[width=1\linewidth, clip]{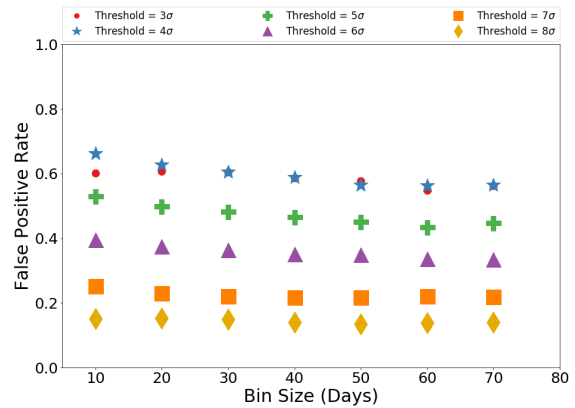}}
    \caption{ As Figure 4, but showing the False Positive Rate (FPR), the fraction of non-microlensing events that are labeled as microlensing. The value of FPR is only weakly dependent of the bin sizes for most thresholds. }
    \label{fig:FPR}
\end{figure}

\begin{figure}
    \centerline{\includegraphics[width=1\linewidth, clip]{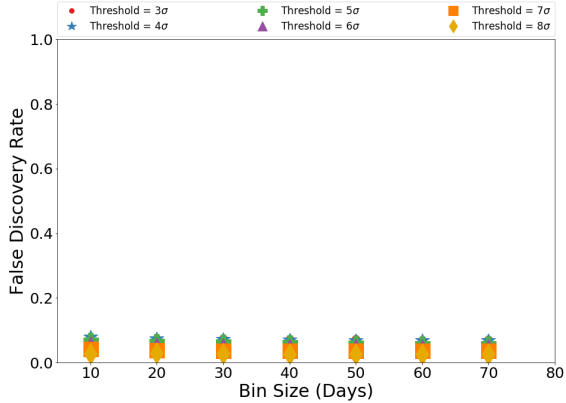}}
    \caption{ As figure 4, but showing the False Discovery Rate (FDR), the fraction of events labeled as microlensing that are in fact non-microlensing. Overall, values of FDR are very low and also roughly independent of the bin sizes for most thresholds.}
    \label{fig:FDR}
\end{figure}

Figures \ref{fig:TPR}, \ref{fig:FPR}, and \ref{fig:FDR} show the performance of the algorithm across a range of bin sizes and thresholds. They display the True Positive Rate (TPR), False Positive Rate (FPR), and False Discovery Rate (FDR) for different options of bin size and peak threshold. Figure \ref{fig:TPR} shows that bin sizes larger than about 50 days and a thresholds of $3\sigma$ to $5\sigma$ give the highest TPR of about 80\%.

Figure \ref{fig:FPR} shows the False Positive Rate which represents what fraction of non-microlensing events are labeled as microlensing.  Bin sizes do not consistently affect the FPR. Thresholds larger than $5\sigma$ give unreasonably low FPR, and that is because at these thresholds, most light curves will have zero peaks identified. Thresholds of $3\sigma$ to $5\sigma$ give a FPR of 50\%. Since some fraction of non-microlensing events have a single peak, we expect that a group of non-microlensing events will be classified as microlensing even though the algorithm has successfully identified one peak in them.

Figure \ref{fig:FDR} show the False Discovery Rate which represents the fraction of events labeled as microlensing that are are in fact non-microlensing. For the same reason as mentioned above, thresholds larger than $5\sigma$ are not favorable, and thresholds of $3\sigma$ to $5\sigma$ have reasonably low FDR of larger than 7\% for most bin sizes. Note that the extremely low rates of FDR is also an indication of the unbalanced dataset (a dataset in which there is unequal number of objects in each category.), and our purpose is to compare how sensitive it is to different thresholds and bin sizes.

Using this peak finder algorithm with a bin size of $60$ days and a threshold of $5\sigma$, we can identify which light curves show a single peak event.  In Section \ref{sec:Approach}, we use this feature in conjunction with the other feature detection tools.

\subsection{Gould Two-parameter PSPL Fit} 
\label{subsec:gpspl}

The PSPL function has been demonstrated to be a good fit to most microlensing light curves, and to not match most other types of astrophysical variability \citep{di1997identifying}. Therefore, the goodness of fit to the PSPL model can be used as a feature of the light curve. In this section, we demonstrate employing a two-parameter PSPL model first introduced by \citet{gould1996theory} to detect curves similar to a Paczy\'nski curve. The reason for not using the regular PSPL model is that brute force searches over the three non-linear parameters ($t_E$, $t_0$, $u_0$) of a PSPL model are slow, while the two-parameter Gould approximate PSPL model (G-PSPL) captures the basic observables of a full PSPL model but requires many fewer computations.  For more details refer to \citet{gould1996theory} and \citet{wozniak1997microlensing}.

In order to employ the PSPL model as a light curve feature, we employ a version of the \citet{gould1996theory} two-parameter PSPL model (G-PSPL) that has been used by the KMTNet survey to detect microlensing events. For that survey, \citet{kim2018korea} fit a combination of two two-parameter G-PSPL models: one with the assumption of $u_0 = 1$, and one with the assumption of $u_0$ being very close to zero. This approach reduces the number of fitted parameters, and can be used to detect both high and low-magnification events. This double two-parameter G-PSPL fit is defined here as function $F(Q)$:
\begin{equation}\label{eqn:F_Gould_2par_PSPL}
    F(Q) = f_1 \times (A1(Q)+A_2(Q)) + f_0
\end{equation}
where
\begin{equation}\label{eqn:A1_Gould_2par_PSPL}
    A_1(Q) = \frac{1}{\sqrt{Q}}
\end{equation}
and
\begin{equation}\label{eqn:A2_Gould_2par_PSPL}
    A_2(Q) = \frac{1}{\sqrt{1-\frac{1}{(1+\frac{Q}{2})^2}}}
\end{equation}
are functions that represent good fits to high and low magnification events, respectively. The two parameters $f_0$ and $f_1$ do not have a physical meaning in this equation, but in the limit of $u_0=1$ and $u_0$ close to zero, they are related to the blending and source parameters, described in detail in \citet{kim2018korea}. $A_1$ and $A_2$ are both functions of $Q$,
\begin{equation}\label{eqn:Q}
    Q(t) = 1 + \left(\frac{t-t_0}{t_{eff}}\right)^2.
\end{equation}
$Q$ is a function of time and depends on two parameters $t_0$ and $t_{\rm eff}$. $t_0$ is the time of the maximum magnification, and $t_{\rm eff}$ approximates $u_0 \times t_E$ in the limit of $u0\ll 0.5$, and helps characterize the amplitude and duration of the microlensing event. For fitting the function $F(Q)$, we set the initial values of the parameters $f_0$ and $f_1$ to $0.5$, and $t_0$ is the time of the maximum magnification in the light curve. For $t_{\rm eff}$, we choose a list of seven initial values $0.01, 0.1, 0.5, 2, 5, 10,$ and $20$. For high magnification events, low values of $t_{\rm eff}$ usually provide a better fit, and for low magnification events, larger values of $t_{\rm eff}$ are a better fit. The fit with the minimum ${\chi}^2$ value is used as the selected model. Note that for the remainder of the paper, we specify whether we use A G-PSPL or a PSPL model fit.

\subsection{Symmetry Check Algorithm}\label{subsec:symcheck}

If we identify a peak in a light curve and fit a two-parameter G-PSPL model to the data, we then employ a check on the symmetry of the event. In Section \ref{subsec:gpspl}, we showed that this method can be used to detect microlensing events. Here, we calculate the reduced ${\chi}^2$ statistic separately for the right and left side of the peak in the light curve. If the peak is symmetric, the ratio of ${{({\chi}^2_{left})}_{reduced}}/{{({\chi}^2_{right})}_{reduced}}$, which we define as $\beta$, is very close to unity. Since the number of data points in the right and left wings of the peak might not be exactly equal, we use the reduced ${\chi}^2$. Deviations of $\beta$ from unity can be related to additional physical details in the light curves, such as the existence of planetary or binary star companion lensing signatures, parallax effects, or the asymmetry in non-microlensing peaks like cataclysmic variables.
 
\begin{figure}
    \centerline{\includegraphics[width=1\linewidth, clip]{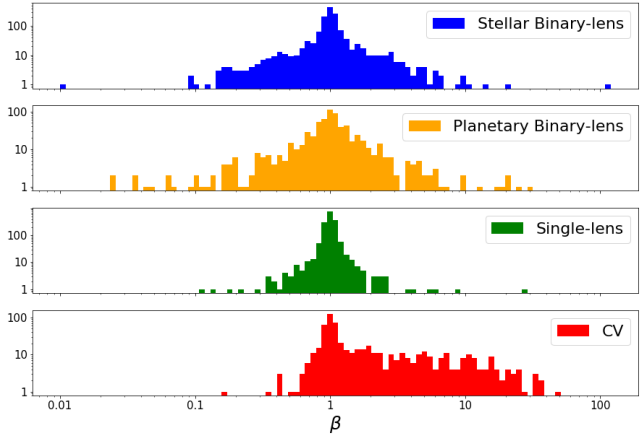}}
    \caption{$\beta$ is a measurement of asymmetry of the peaks in the light curves that is obtained from the symmetry check algorithm. This plot shows the distribution of $\beta$ for different classes of light curves in our dataset.}
    \label{fig:Gould_2_par_PSPL_confusion_matrix}
\end{figure}

In Figure \ref{fig:Gould_2_par_PSPL_confusion_matrix} we plot histograms showing the distributions of variability types for a set of ranges of $\beta$ values. Each column represents events with a specific range of $\beta$. We find that for CV light curves, the asymmetry is extremely large, consistent with the morphology of CV light curves, displaying steep rises and more gradual decreases.  Single-lens microlensing events are symmetric, with $\beta$ values close to unity. Binary lenses have a wide range of $\beta$ values depending on whether they are caused by a planetary system or a stellar binary system. These ratios can be used as a feature to obtain information about the kinds of deviations from the PSPL model, and classify the light curves.
 
\subsection{Trapezoidal Function Fit} \label{subsec:trap}

As explained in Section \ref{subsec:free}, lensing events caused by free-floating planets have a short duration and amplitude and are more often strongly affected by the finite source effect. Therefore, their shapes approximately resemble a trapezoidal function. 

\begin{figure}
    \centerline{\includegraphics[width=1\linewidth, clip]{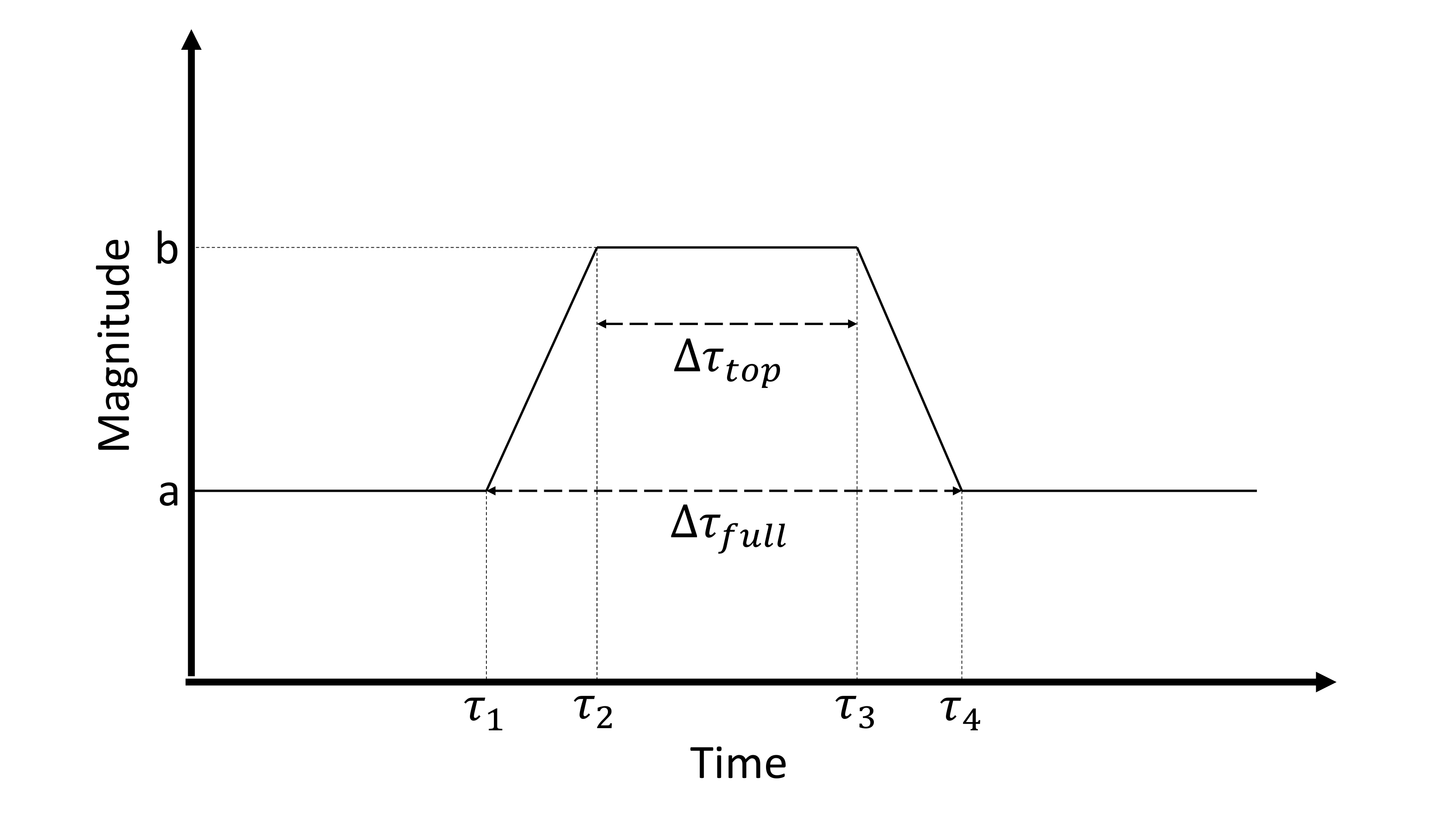}}
    \caption{A trapezoidal function with its defined parameters.}
    \label{fig:trapezoid_params}
\end{figure}

The parameters of the trapezoidal function are the baseline magnitude ($a$), maximum magnitude ($b$), time of the first rise (${\tau}_1$), the time when maximum magnitude is reached (${\tau}_2$), the time when maximum magnitude ends (${\tau}_3$), and the time when the magnitude returns to the baseline (${\tau}_4$). Figure \ref{fig:trapezoid_params} shows a diagram of the function and its six parameters. We must first find initial guesses for the six parameters, then fit the function and obtain more accurate estimates of the parameters. We also calculate the duration of the trapezoidal portion $\Delta \tau_{full}$, and the duration of the flat section of the trapezoid, $\Delta \tau_{top}$ as presented in Equations \ref{eqn:tau_full} and \ref{eqn:tau_top} and shown on Figure \ref{fig:trapezoid_params}.

\begin{equation}\label{eqn:tau_full}
    \Delta \tau_{full} = {\tau}_4 - {\tau}_1
\end{equation}

\begin{equation}\label{eqn:tau_top}
    \Delta \tau_{top} = {\tau}_2 - {\tau}_3
\end{equation}

In order to find initial guesses for the six parameters, we first subtract the median baseline magnitude from the light curve, setting $a$ to zero. We then select $20$, $2$, $0.2$, or $0.02$ days as initial guesses for the total duration ($\Delta \tau_{full}$).  We define $t_0$ as the time of maximum brightness, and assume $\Delta \tau_{full} = 2\Delta \tau_{top}$.  We then define the remaining parameters ${\tau}_1$, ${\tau}_2$, ${\tau}_3$, and ${\tau}_4$ as the following quantities, respectively: $t_0 - \frac{\Delta \tau_{full}}{2}$, $t_0 - \frac{\Delta \tau_{full}}{4}$, $t_0 + \frac{\Delta \tau_{full}}{4}$, and $t_0 + \frac{\Delta \tau_{full}}{2}$. We fit the resulting models based on the initial guesses for $\Delta \tau_{full}$, using a least-squared minimization approach.  We finally select the fit that minimizes the ${\chi}^2$ statistic.

\begin{figure}
    \centerline{\includegraphics[width=1\linewidth, clip]{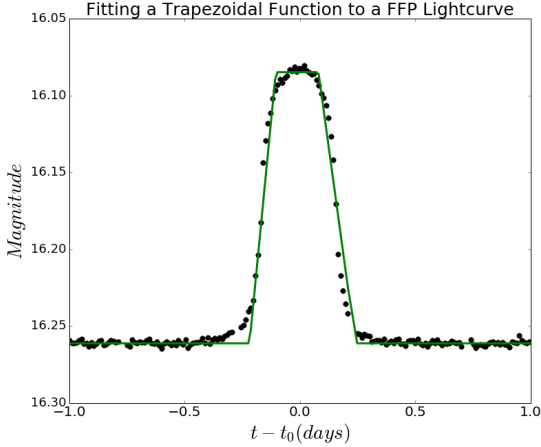}}
    \caption{Trapezoidal function fitted to the light curve of a simulated event caused by a free-floating planet. The flat top of the trapezoid describes the duration of maximum brightness fairly well, and the duration of the event is well represented by the full trapezoidal duration.}
    \label{fig:Trapezoid_example}
\end{figure}

\begin{figure}
    \centerline{\includegraphics[width=1\linewidth, clip]{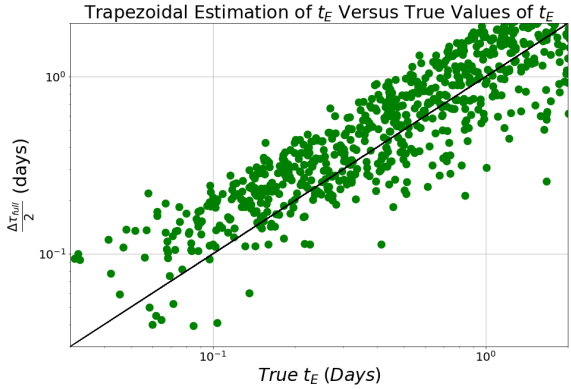}}
    \caption{True versus estimated Einstein Timescale ($t_E$) of the events with $t_E<1$ days found by the trapezoidal function fit. These events are candidates for free-floating planets.}
    \label{fig:tE_FFP_trapezoid}
\end{figure}

We have applied this algorithm to our test dataset, with Figure \ref{fig:Trapezoid_example} showing an example of a best fit function for a simulated light curve.  We find that half of the total duration ($\Delta \tau_{full}$) is a good representation of the Einstein timescale ($t_E$) of the event, and we compare it with true values of $t_E$ of the events. Figure \ref{fig:tE_FFP_trapezoid} shows the estimated versus true duration of single-lens microlensing events with $t_E < 2$ days that are candidates for free-floating planets. We find that for events with $t_E < 2$ days, there is a median absolute deviation of $0.06$ days, while for events with $t_E \ge 2$ days, there is a median absolute deviation of $0.8$ days.

We define another parameter $\kappa = \frac{\Delta \tau_{top}}{\Delta \tau_{full}}$ as the trapezoidal timescale ratio, which is the ratio of the duration of the flat part of the trapezoid to the total duration of the trapezoid. This quantity was initially set to 0.5 for the initial guesses of the function fitting.  This parameter can be thought of as a measure of how square or peaked the event is, analogous to the kurtosis of the event. We show that using $\kappa$ along with $\Delta \tau_{full}$ allows us to flag the events that are strongly affected by the finite source effect.  We can characterize the approximate significance of the finite source effect using $|u_0|/\rho$, the ratio of the impact parameter to the finite source ratio.  When $|u_0|/\rho$ is smaller than unity, it implies the lens passes directly over the source and thus that can be a sign of significant finite source effects.

\begin{figure*}[t]
    \centerline{\includegraphics[scale=0.4, clip]{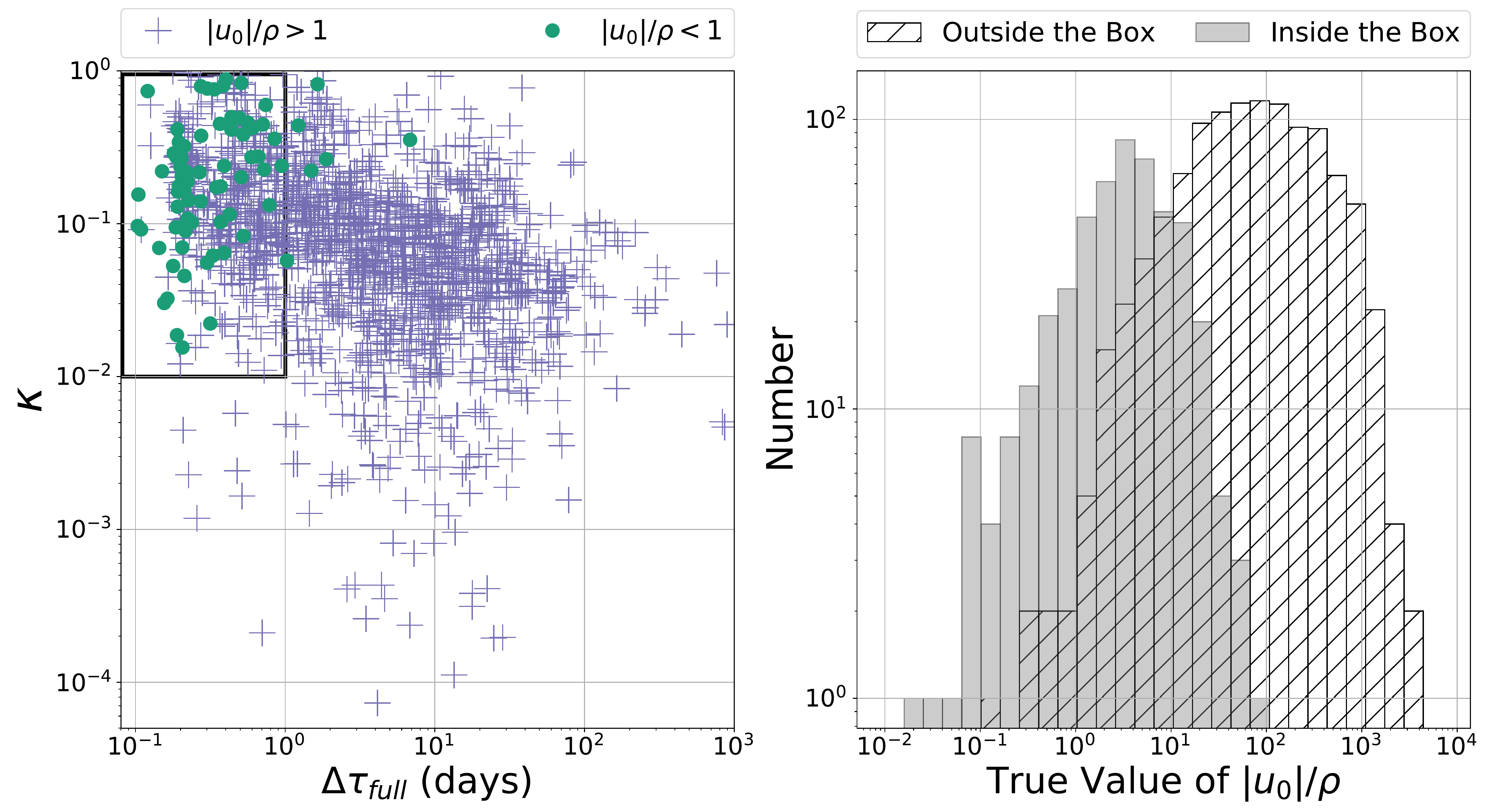}}
    \caption{ {\it Left Panel}: Values of the trapezoidal timescale ratio ($\kappa$) are plotted versus the total trapezoidal duration ($\Delta \tau_{full}$). The purple data points show events with $|u_0|/\rho>1$, and the green data points represent events with $|u_0|/\rho<1$, which indicates events with free-floating planets or other PSPL events affected by the finite source effect. $\kappa$ is larger for shorter events, and most of the events with $|u_0|/\rho<1$ have large $\kappa$ and small $\Delta \tau_{full}$. Therefore, the selected ranges of $\kappa$ and $\Delta \tau_{full}$ can be used to identify likely events caused by free-floating planets. {\it Right Panel}: The distribution of $|u_0|/\rho$ for all data points inside (grey) and outside (hatched white) of the black box in the left panel. Light curves with small values of $\Delta \tau_{full}$ and large values of $\kappa$ tend to have smaller values of $|u_0|/\rho$, and light curves with larger values of $\Delta \tau_{full}$ and smaller values of $\kappa$ tend to have larger values of $|u_0|/\rho$.} 
    \label{fig:rho_of_tE_less_one}
\end{figure*}

The left panel of Figure \ref{fig:rho_of_tE_less_one} shows the trapezoidal timescale ratio ($\kappa$) versus $\Delta \tau_{full}$ for the subset of all light curves that meet the fitting criteria for single-lens models, which are described below in \S \ref{sec:Approach}. Note that both of the quantities on the axes in the left panel are found experimentally using the trapezoidal function fit.  In that panel, most of the green data points have larger values of $\kappa$ which implies a more square light curve shape caused by strong finite source effects. The right panel shows the distributions of the true values of $\rho$ for all the data points inside and outside of the black box in the left panel. That panel indicates that light curves with small values of $\Delta \tau_{full}$ and large values of $\kappa$ tend to have larger values of $\rho$ (grey), and light curves with larger values of $\Delta \tau_{full}$ and smaller values of $\kappa$ tend to have smaller values of $\rho$ (hatched white). This is because events with short duration caused by free-floating planets are more affected by the finite source effect, and therefore more resemble a trapezoidal functions, with $\Delta \tau_{full}$ and $\kappa$ helping to indicate these events.

\subsection{Cauchy Distribution Fit}\label{subsec:cauchy}

For microlensing events with a strong finite source effect in the absence of limb darkening effects, the top of the peak becomes flatter and will not look like a PSPL function. For detecting this phenomenon, we need to parameterize the flatness of the top of the peak. Effectively, we would like to use a functional fit that can apply to the range of morphologies between that of a single-lens PSPL model, and a trapezoid that much more closely fits a strong finite source effect.  Our goal here is to find a correlation between the fitted parameters and the finite source ratios. For that purpose, we use a Cauchy distribution function to fit the light curves.  The difference in the minimum ${\chi}^2$ of the Cauchy fit and the PSPL fit (shown in Equations \ref{eqn:PSPL_A_fs} and \ref{eqn:A_PSPL}), along with one of the parameters of the Cauchy distribution, can be used as features.

The Cauchy distribution shown in Equation \ref{eqn:Bell} has four parameters, time of the peak ($t_0$), the duration of the peak ($\sigma$), the flatness of the peak ($b$), and the amplitude of the peak ($a$).
\begin{equation}\label{eqn:Bell}
C(t) = \frac{a}{1+{\left|\frac{t-t_0}{\sigma}\right|}^{2b}}
\end{equation}
The functions $C(t)$ and PSPL have two common parameters, time and duration of the peak. After fitting these two functions to a light curve, we have fitted values for $t_E$ and $t_0$, which we refer to as  $t_{E,PSPL}$, $t_{E,Cauchy}$, $t_{0,PSPL}$, and $t_{0,Cauchy}$. Note that $t_{E,Cauchy}$ is equivalent to $\sigma$ in Equation \ref{eqn:Bell}. We define the difference in the ${\chi}^2$ of the PSPL fit and the Cauchy fit as a feature characterizing the flatness of the top of the curve represented in Equation \ref{eqn:psi}. In order to calculate $\psi$, we select a section of the curve at the peak, between $t_0 - (t_{E,PSPL} \times u_{0,PSPL})$ and $t_0 + (t_{E,PSPL} \times u_{0,PSPL})$.

\begin{equation}\label{eqn:psi}
    \psi =  {{\chi}^2}_{PSPL}-{{\chi}^2}_{Cauchy}
\end{equation}

\begin{figure}
        \centerline{\includegraphics[width=1\linewidth, clip]{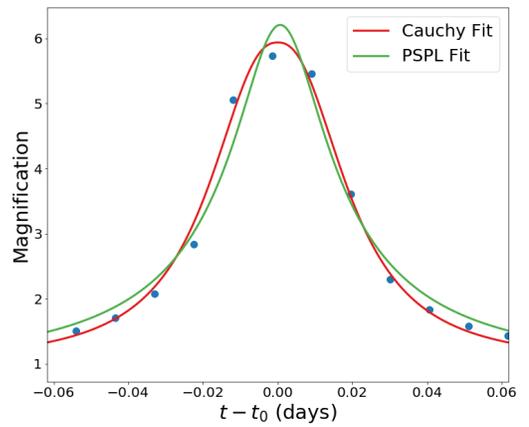}}
        \caption{Cauchy and PSPL functions fitted to a Roman simulated single-lens microlensing light curve highly affected by the finite source effect. The blue data points are from the simulated light curve, the red curve is the fitted PSPL model, and the orange curve is the fitted Cauchy model. }
        \label{fig:Bellcurve}
    \end{figure}

The more the event is affected by the finite source effect, the more the top of its peak deviates from the PSPL model and is more similar to the Cauchy model. Thus for single-lens events with a flatter peak, $\psi$ is positive. Even in cases where the Cauchy distribution fits much better than the PSPL fit, for events that are not caused by free-floating planets that still experience strong finite source effects, the Cauchy distribution will typically be a better fit than the trapezoidal function. Figure \ref{fig:Bellcurve} shows a single-lens event highly affected by the finite source effect, along with the best-fit PSPL and Cauchy models. The two curves deviate the most close to the peak of the event.

\begin{figure*}[!t]
    \centerline{\includegraphics[scale=0.6, clip]{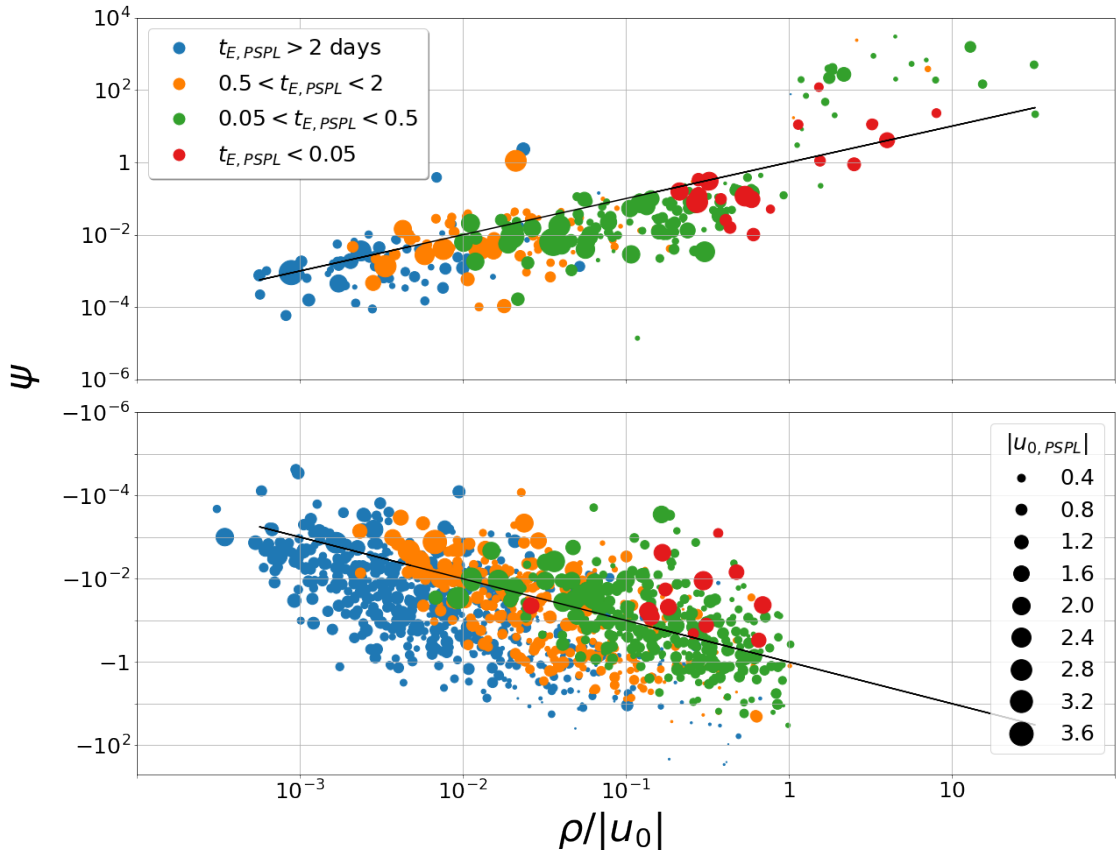}}
    \caption{ The Cauchy feature ($\psi$) plotted versus $\rho/u_0$, restricted to cases with at least four observations near the peak maximum. The colors of the points represent the PSPL-estimated event duration $t_{E,PSPL}$, while the sizes of the points represent the PSPL-estimated value of $u_0$ ($u_{0,PSPL}$).
    The black trend lines show a one-to-one relation between $\psi$ and $\rho/u_0$ (reversed in the bottom panel). }
    \label{fig:Cauchy_delta_chi_rho}
\end{figure*}

Figure \ref{fig:Cauchy_delta_chi_rho} shows values of the Cauchy feature ($\psi$) versus $\rho/u_0$ for all single-lens events that have at least four observations within a time of $t_{E,PSPL} \times u_{0,PSPL}$ from the maximum. The upper panel shows positive values of $\psi$ and the lower panel shows the negative values.  The events with flatter peaks have a positive value of $\psi$, and on average, have a shorter duration than those with negative $\psi$. Events with large positive $\psi$ (flatter peaks) appear in the upper right of this figure and have $\rho/u_0>1$.  Events with a negative $\psi$ include both short and long duration events. The black lines show the one-to-one relation for events with positive and negative $\psi$, and have a median absolute deviations of 0.45 for positive $\psi$, and 0.64 for negative $\psi$. Events in the lower region mostly have small negative values of $\psi$, indicating that neither the PSPL and Cauchy functions are significantly better fits to the data.  Those with large negative values are those with sharp peaks that are well-described by the PSPL function, which also have low $u_0$. This plot shows that the sign and magnitude of $\psi$ can help identify the flattest and sharpest microlensing peaks, and flag the ones that are more likely affected by the finite source effect.

\begin{figure}
    \centerline{\includegraphics[width=1\linewidth, clip]{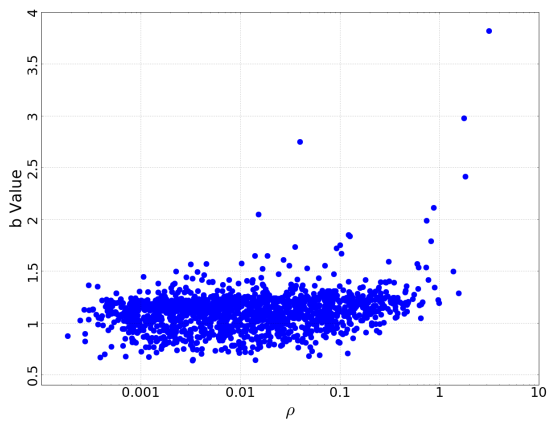}}
    \caption{Plot of the parameter $b$ in equation \ref{eqn:Bell} versus true parameters of $\rho$. It shows that the $b$ values are in general larger for events with $\rho$ larger than $0.1$.}
    \label{fig:Cauchy_b_vs_rho}
\end{figure}

The other feature that we use for this purpose is parameter $b$ in Equation \ref{eqn:Bell}. We expect that if the event exhibits strong finite source effects which means a flatter top, this feature would take a larger value, and if the finite source effect is negligible, the feature would be very close to unity (in the absence of limb darkening effects). In reality, $b$ does not correlate with $\rho$ for most events, but once $\rho$ is greater than 0.1, we do see a positive correlation with $b$, as seen in Figure \ref{fig:Cauchy_b_vs_rho}.  We therefore retain $b$ as a potentially useful feature.

\subsection{Planetary Parameters Finder} \label{subsec:busy}

Our goal in this section is to measure the two physical parameters associated with the planetary binary-lens events. These are $s$, the projected separation between the planet and the lens star and $q$, the mass ratio of the planet and the lens star. The approach that we take here is to first fit the main event with a PSPL function, then find the deviations from the PSPL model by looking at the residual, to identify peaks or troughs. We then seek to identify cases where these residuals have one significant peak or trough, or if they have two significant peaks. 

The reason we take this approach is that the single-deviation residuals can be characterized much more robustly than the double-peaked ones. For events of the both groups, we fit the ``busy" function introduced by \citet{westmeier2014busy}. This function is primarily used to describe double-peaked features in spectra, and with some changes in its parameters, we can also use it to find the single-peaked deviations. \citet{khakpash2019fast} suggest fitting the single-peaked residuals with a Gaussian, and show that this approach is useful mostly for low-mass ratio events. Here, we take the same approach to find the initial parameters in the residual, but instead of a Gaussian, we fit all deviations with the busy function. After fitting the residual, in the next round of fittings, we fit a PSPL plus a busy function, and we use parameters obtained from the first two fits as initial parameters of this fit. Next, we calculate $s$ and $q$ using the parameters found by the final fit. 

\begin{figure}
    \centerline{\includegraphics[width=1\linewidth, clip]{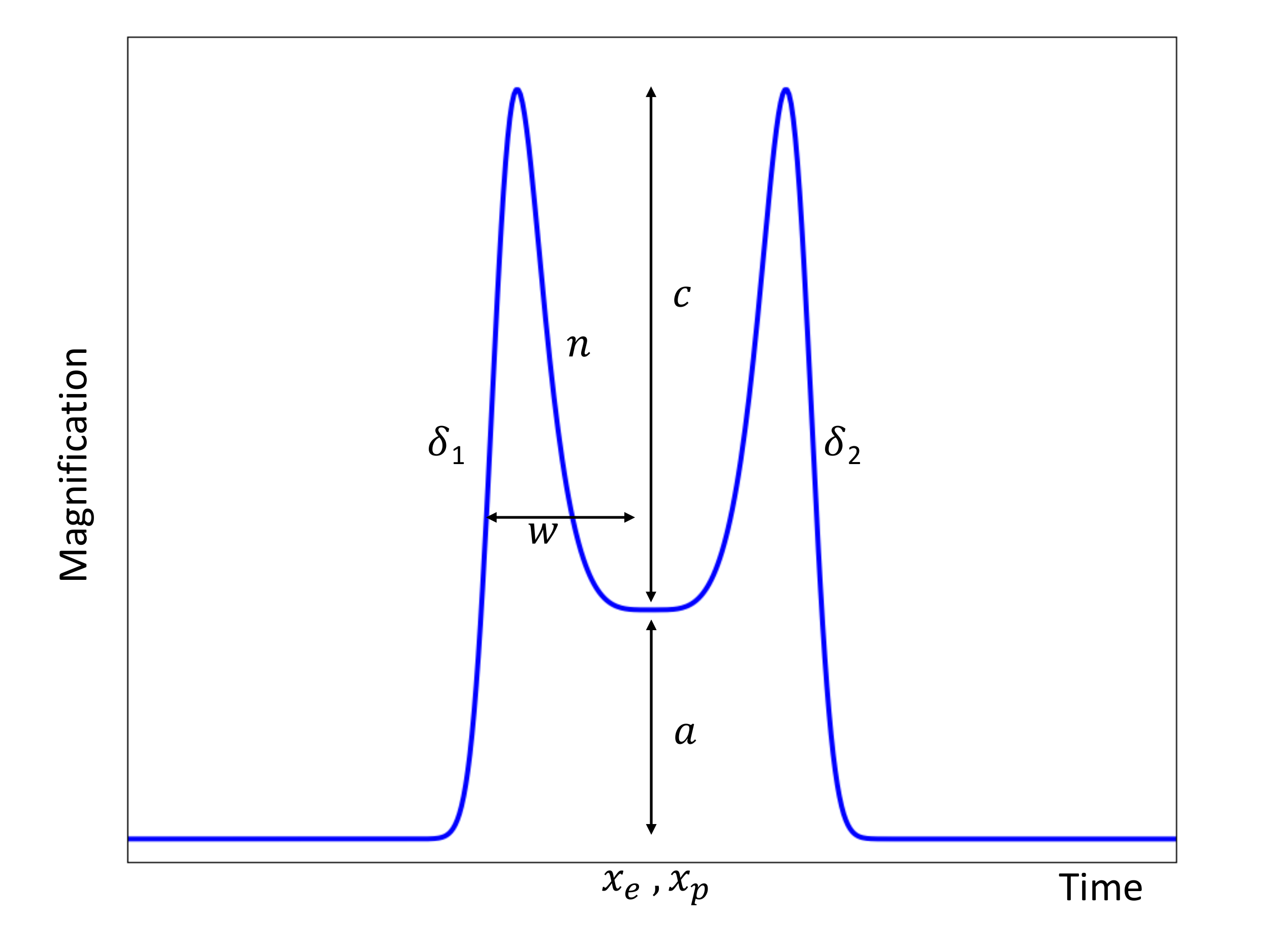}}
    \caption{A schematic plot of the busy function, with features associated with eight function parameters. The parameter $\varepsilon$ (not shown on the figure) represents a horizontal scaling of the function without disrupting its shape.}
    \label{fig:Double_horn_params}
\end{figure}

The busy function is commonly used to describe double-horn profile of galaxy spectra \citep{westmeier2014busy}. The function is shown in Equation \ref{eqn:double_horn}, and has nine parameters that are sketched in Figure \ref{fig:Double_horn_params}. This function comprises two error functions and a polynomial of degree $n$. The parameters $x_e$ and $x_p$ determine location of the middle of the two error functions and the middle of the polynomial, ${\delta}_1$ and ${\delta}_2$ are the steepness of the two error functions, $n$ is the degree of the polynomial and determines the steepness of the middle trough, $c$ determines the depth of the polynomial, $w$ determines the distance of the error function zero points from $x_e$, $a$ determines the height of the error functions, and $\varepsilon$ is the horizontal scaling of the function. 
    
\begin{equation}\label{eqn:double_horn}
\begin{split}
A(t) \; =\; (a/4)\; \times \; \; ({\rm erf}\; (\;{\delta}_1 (\;w + \varepsilon \times t- x_e)) + 1)\; \\
\times \; \; ({\rm erf}\; (\;{\delta}_2(\;w - \varepsilon \times t + x_e)) + 1)\; \\
\times \; \; (c\; \times \; {|\varepsilon \times t-x_p|}^n+1)
\end{split}
\end{equation}

\begin{figure*}[t]
          \centerline{\includegraphics[scale=0.35, clip]{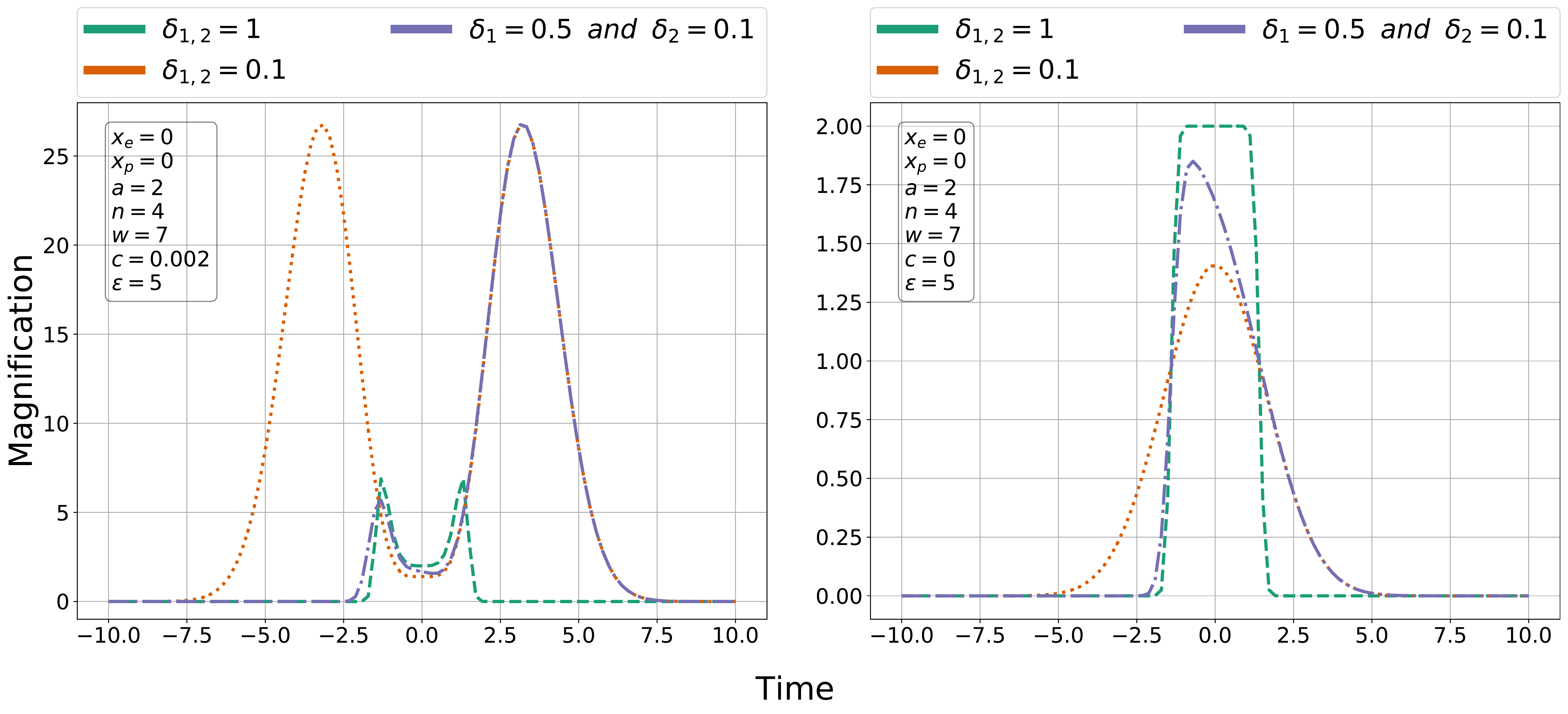}}
          \caption{Two different shapes of the busy function that are fitted to the single- and double-peaked residuals. {\it Left Panel}: A double-horn shape of the busy function, and the values of the parameters. The three curves represent the shapes of the function when only the steepness of the two error functions are altered. {\it Right Panel}: The value of $c$ in Equation \ref{eqn:double_horn} is set to zero, and the equation only comprises the two error functions and no polynomial. Only one peak is present, with the two error functions controlling the steepness of the two sides of the peak. The three curves show how the shape of the function changes as the steepness of the two error functions are altered.}
          \label{fig:Double_horn_forms}
        \end{figure*}

Depending on the values of the nine parameters, the shape of the function can differ significantly. We take advantage of this fact and use different shapes of the function to fit single-peaked and double-peaked deviations from the PSPL model. The two forms of the busy function used to fit the residuals are shown in Figure \ref{fig:Double_horn_forms}. The left panel shows a double-horn shape of the busy function, and the values of the parameters can be found on the plot. The three curves represent the shapes of the function when only the steepness of the two error functions are altered. This set of parameters results in a form that can describe the caustic-crossing features of the planetary microlensing light curves. In the right panel, the value of $c$ in Equation \ref{eqn:double_horn} is set to zero, and therefore the equation only comprises two error functions. The shape of the function resembles the Gaussian function used by \citet{khakpash2019fast} to fit single-peaked deviations with an additional ability to describe asymmetric deviations. The three curves show how the shape of the function changes as the steepness of the two error functions are altered.
        
We apply this approach to the set of simulated Roman planetary microlensing light curves. We follow the procedure in \citet{khakpash2019fast}, and first fit a PSPL function as shown in Equations \ref{eqn:A_PSPL} and \ref{eqn:PSPL_A_fs} to the light curves. We find an initial value for $t_0$ by finding the time of the maximum flux in the light curve, and we set the initial value of $f_s$ to be $0.5$. For an initial guess for the duration of the event $t_E$, we first interpolate between the data points using a cubic interpolation to estimate a continuous version of the data, then, for events with $u_{0,initial}<0.5$, we assume $t_E$ is the interval between times when the magnification is $1.34$. For events with $u_{0,initial}>0.5$, we set $t_E$ equal to the intervals between times when magnification is $1.06$.  We use Equations \ref{eqn:A_PSPL} and \ref{eqn:PSPL_A_fs} to calculate an initial guess for $u_0$ based on the above other assumed values at time $t=t_0$. Then, we fit the PSPL function using the initial guesses, and then apply the peak-finding algorithm to the residual.

Next, we use the peak finder (\S \ref{subsec:PeakFinder}) to look for deviations in the residual, and we force it to find one or two peaks. We start with a large bin size of about half the lightcurve baseline and a threshold of $3\sigma$. With these values, the algorithm finds the most significant deviation. Then, we start decreasing the bin size by 50\% until it either finds two peaks or the bin size reaches 0.2 days. At any of these decreasing steps, if it finds two peaks that are separated by less than 10 days, we end the search. This condition aims at preventing the algorithm from selecting multiple peaks in a caustic-crossing microlensing event. If the search concludes with the detection of one deviation, we accept that and move on to the next step. If it finds zero deviations, we simply select the maximum or minimum data point in the residual and identify that as a single deviation. 

At this point, we fit the modified busy function to peaks identified in the residual. For this purpose, we identify the time of the single deviation, or the middle point of the double-peaked deviation to be initial guesses for $x_e$ and $x_p$. We set these two values equal to zero, the values of ${\delta}_1$ and ${\delta}_2$ are set equal to $1$, and the value of $\varepsilon$ is set to $5$. Parameters $a$, $w$, $n$ and $c$ are set up differently depending on the number of deviations found in the residual. If there is one deviation, $n$ and $c$ are set equal to zero, and $a$ is equal to the largest peak or trough in the residual, and $w$ is set equal to unity. If there are two deviations, initial values of $n$ and $c$ are set to $10$ and $0.02$, and the value of $a$ is set equal to the median of the residual values between the two peaks, and $w$ is set equal to half of the distance between the two peaks.

Parameters describing the deviations in the residual are obtained by the busy function fit. Using these parameters along with the PSPL parameters as initial values, we then fit a PSPL plus a busy function to the light curve, and we find final values of $13$ parameters, four of them describing the main event, and nine of them describing the deviation. Using these values, we calculate the two physical parameters $s$ and $q$ using two different approaches.

When only one deviation is found, the busy function will turn into two error functions with six parameters. After fitting the function to the deviation, we determine the duration of the deviation by looking at where the fitted model is not zero, and we assume half of that to be the duration $t_{Ep}$. $t_p$ is set to $x_e$, and then we find the two values of $s$ by solving Equation \ref{eqn:find_s}. If the deviation is positive, a secondary check is done to find if there is a large trough close to this peak. If there is a large de-magnification close to a single caustic crossing, the event is very likely to have $s<1$ which could also be due to a resonant caustic. In the secondary check, ratio of the sum of negative data points in the residual of the busy function fit to the sum of positive data points is calculated, and if the ratio is larger than unity, the value of $s<1$ is chosen, otherwise $s>1$ is selected. If the deviation is negative, the value of $s<1$ is chosen. The estimated value of $q$ is then obtained by Equation \ref{eqn:find_q}.

\begin{equation}\label{eqn:find_s}
    |s - \frac{1}{s}| = u ,\; {\rm where} \; u = \sqrt{{u_0}^2 + {\left(\frac{t_0 - t_p}{t_E}\right)}^2}
\end{equation}

    \begin{equation}\label{eqn:find_q}
        {q} = {\left(\frac{t_{Ep}}{t_{E}}\right)}^2
    \end{equation}

When there are two deviations, we assume that the deviations are caused by crossing the planetary caustic, and that these epochs correspond to when the source approaches two cusps of the caustic. Using these epochs, we calculate the distance of the source from the lens at these times using $u_{1,2} = \sqrt{{u_0}^2 + {\left(\frac{t_0 - t_{1,2}}{t_E}\right)}^2}$. \citet{han2006properties} calculates the size of the planetary caustics in terms of $s$ and $q$, and we find the size of the caustic using simple geometry, and then, we use their formulation to calculate $s$ and $q$.

\begin{figure*}[t]
          \centerline{\includegraphics[scale=0.4]{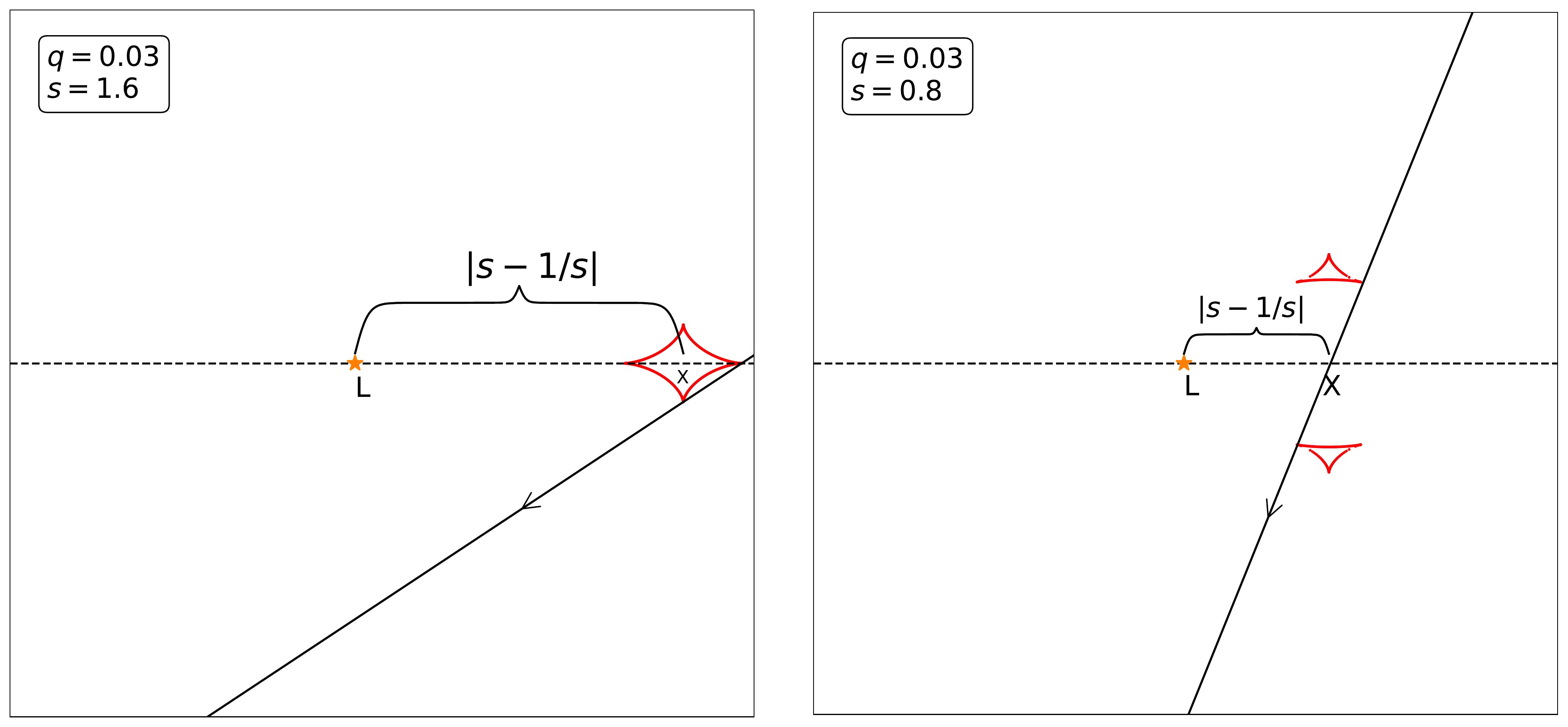}}
          \caption{{\it Left panel}: Planetary caustics of a system with a projected separation of $1.6$ Einstein radius and mass ratio of $0.03$ (red). The size of the caustic varies with $s$ and $q$. {\it Right panel}: Planetary caustics of a system with a projected separation of $0.8$ Einstein radius and mass ratio of $0.03$ (red). At this range of $s$, there are two smaller planetary caustics. The distance between the lens and the center of the planetary caustic is equal to $|s-1/s|$. The solid black line with the arrow in each panel represents the path of the source through the caustics. The left panel is a reproduction of a similar figure in \citet{poleski2014super}.}
          \label{fig:Planetary_Caustics}
        \end{figure*}

Figure \ref{fig:Planetary_Caustics} shows an example of the geometry involved in these lensing events.  In this figure, planetary caustics of two systems with projected separation of $1.6$ and $0.8$ Einstein radius, and mass ratio of $0.03$ is shown. The shape and sizes of the caustics depend on $s$ and $q$. An example path of the source through the system is shown on both panels. The distance between the lens and the center of the caustics, $LX$, gives us the value of $|s- \frac{1}{s}|$. At this point, we do another secondary check as described earlier and we determine whether we have a system with $s>1$ or $s<1$. Next, for systems with $s>1$, we find the vertical dimensions of the caustic (left panel of Figure \ref{fig:Planetary_Caustics}), and for systems with $s<1$, we find the distance between the two caustics by geometry (right panel of Figure \ref{fig:Planetary_Caustics}). We refer to both of these values as height of the caustics. \citet{han2006properties} finds that these values are roughly equal to $\frac{2\sqrt{q}}{s\sqrt{s^2 - 1}}$ and $\frac{s^3  \sqrt{q}}{4}  \sqrt{(\frac{2}{s})^8 + 27}$, respectively. We then use these equations to find $q$. Table \ref{table:Calculate_q_s_process} summarizes the process we adopt to extract the system parameters from the light curve residuals in the cases of one and two deviations \citep{poleski2014super, bozza2000signs}.

\begin{table*}
    \caption{The process taken to calculate $s$ and $q$ is shown in this table.}
    \begin{tabular}{|p{4cm}|p{10cm}|}
        \hline
        \vspace{1mm}
        {\bf One Deviation:}& \begin{minipage}[t]{\linewidth}
        \begin{itemize}
        \vspace{1mm}
        \item[1.] $t_p$ is determined.
        \item[2.] $u = \sqrt{{u_0}^2+{(t_p-t_0/t_E)}^2}$
        \item[3.] $s-\frac{1}{s}-u$ is solved for $s$.
        \item[4.] Secondary check determines if $s>1$ or $s<1$.
        \item[5.] $q={(\frac{t_{E,p}}{t_E})}^2$.\\
        \end{itemize} 
        \end{minipage}\\
     \end{tabular}
        
    \begin{tabular}{|p{4cm}|p{10cm}|}
        \hline
        \vspace{1mm}
        {\bf Two Deviations:} & \begin{minipage}[t]{\linewidth}
        \begin{itemize}
        \vspace{1mm}
        \item[1.] $t_1$ and $t_2$ are determined.
        \item[2.] $u_{1,2} = \sqrt{{u_0}^2+{(t_{1,2}-t_0/t_E)}^2}$
        \item[3.] $LX$ is found by geometry.
        \item[4.] $|s-\frac{1}{s}|-LX$ is solved for $s$.
        \item[5.] Secondary check determines if $s>1$ or $s<1$.
        \item[6.] For $s>1$: Height of the caustic $\sim \frac{2\sqrt{q}}{s\sqrt{s^2 - 1}}\implies q$ is found.
        \item[7.] For $s<1$: Height of the caustic $\sim \frac{s^3  \sqrt{q}}{4}\sqrt{(\frac{2}{s})^8 + 27} \implies q$ is found.\\
        \end{itemize}
        
        \end{minipage}
    \\\hline
    \end{tabular}
\label{table:Calculate_q_s_process}

\end{table*}

\begin{figure}
          \centerline{\includegraphics[width=1\linewidth, clip]{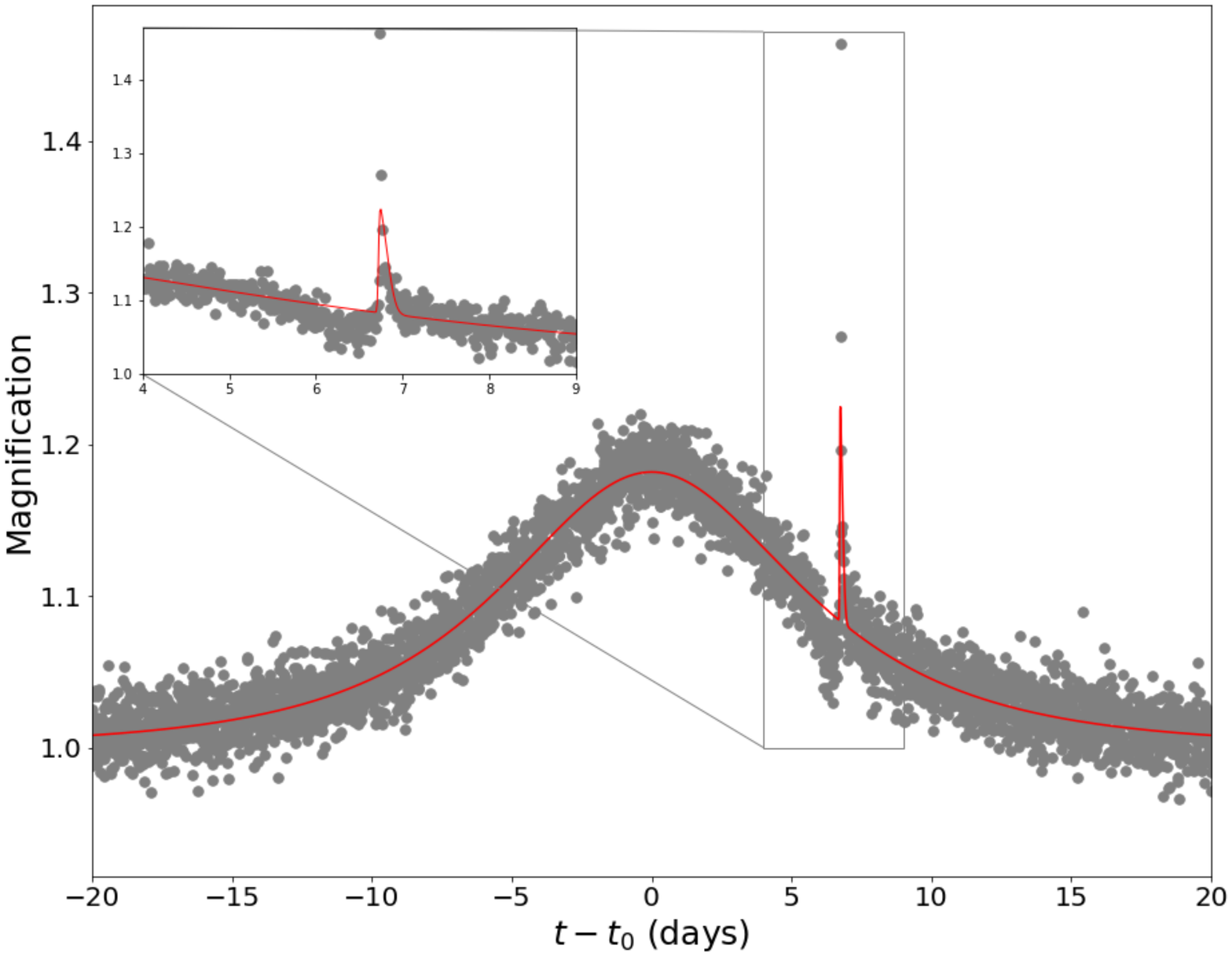}}
          \caption{An example of the PSPL plus the busy function to the light curve of an event with $s=0.63$ and $q=0.0003$. The algorithm detects a sharp peak and would simply decide that the system is caused by a major image if no secondary checks are done. After a secondary check, it detects the negative deviations from the model, and decides that $s$ should be less than unity. The estimated parameters are $s=0.64$ and $q=0.0002$.}
          \label{fig:planetary_ex1}
        \end{figure}

\begin{figure}
          \centerline{\includegraphics[width=1\linewidth, clip]{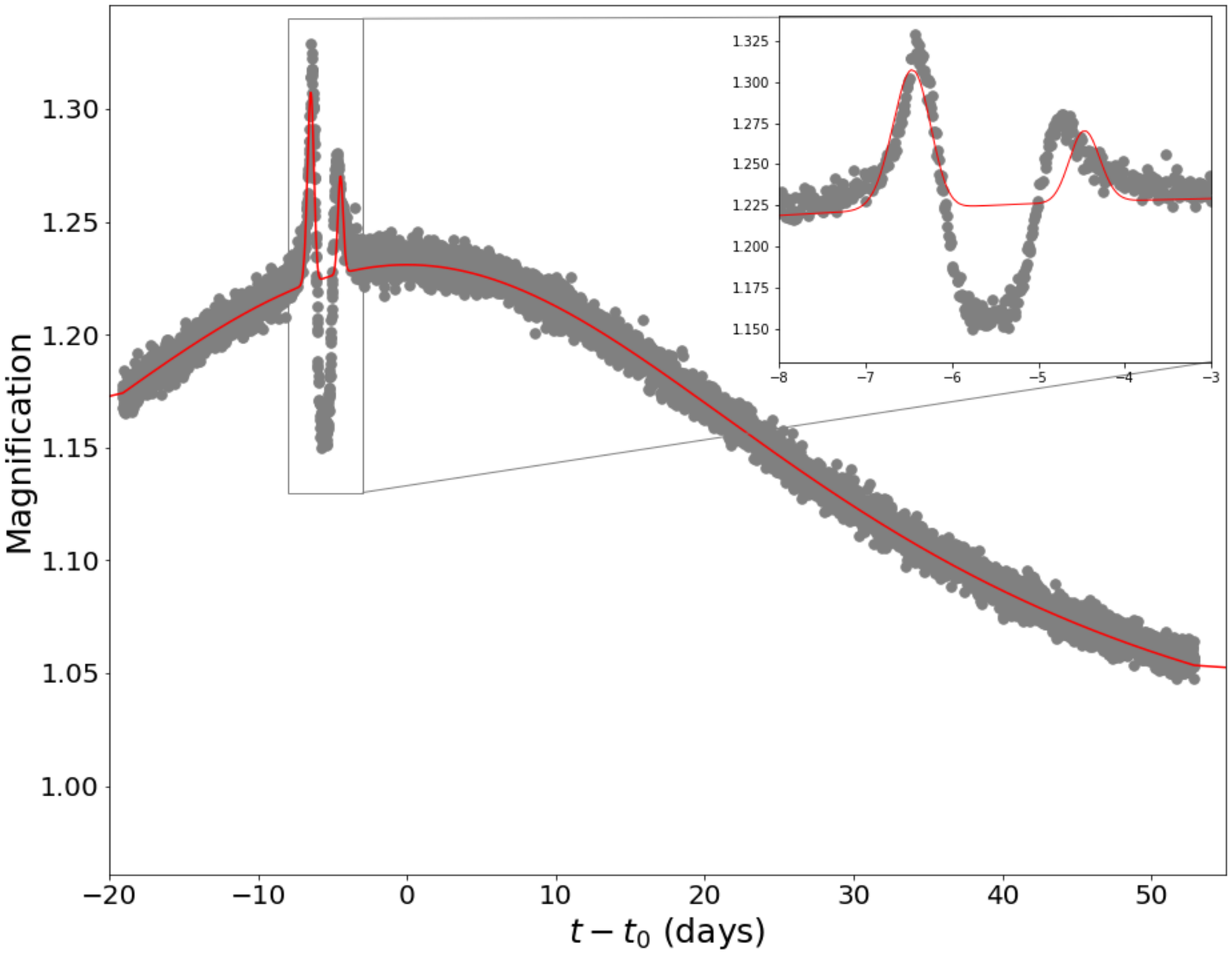}}
          \caption{An example of the PSPL plus the busy function to the light curve of a double-peaked event with $s=0.62$ and $q=0.0001$. The busy function gives us a good estimation of the location and duration of the event, and by doing a secondary check, the large negative deviation from the model is detected, and the algorithm decides that $s$ should be less than unity. The estimated parameters are $s=0.59$ and $q=0.0003$. }
          \label{fig:planetary_ex2}
        \end{figure}

\begin{figure}
          \centerline{\includegraphics[width=1\linewidth, clip]{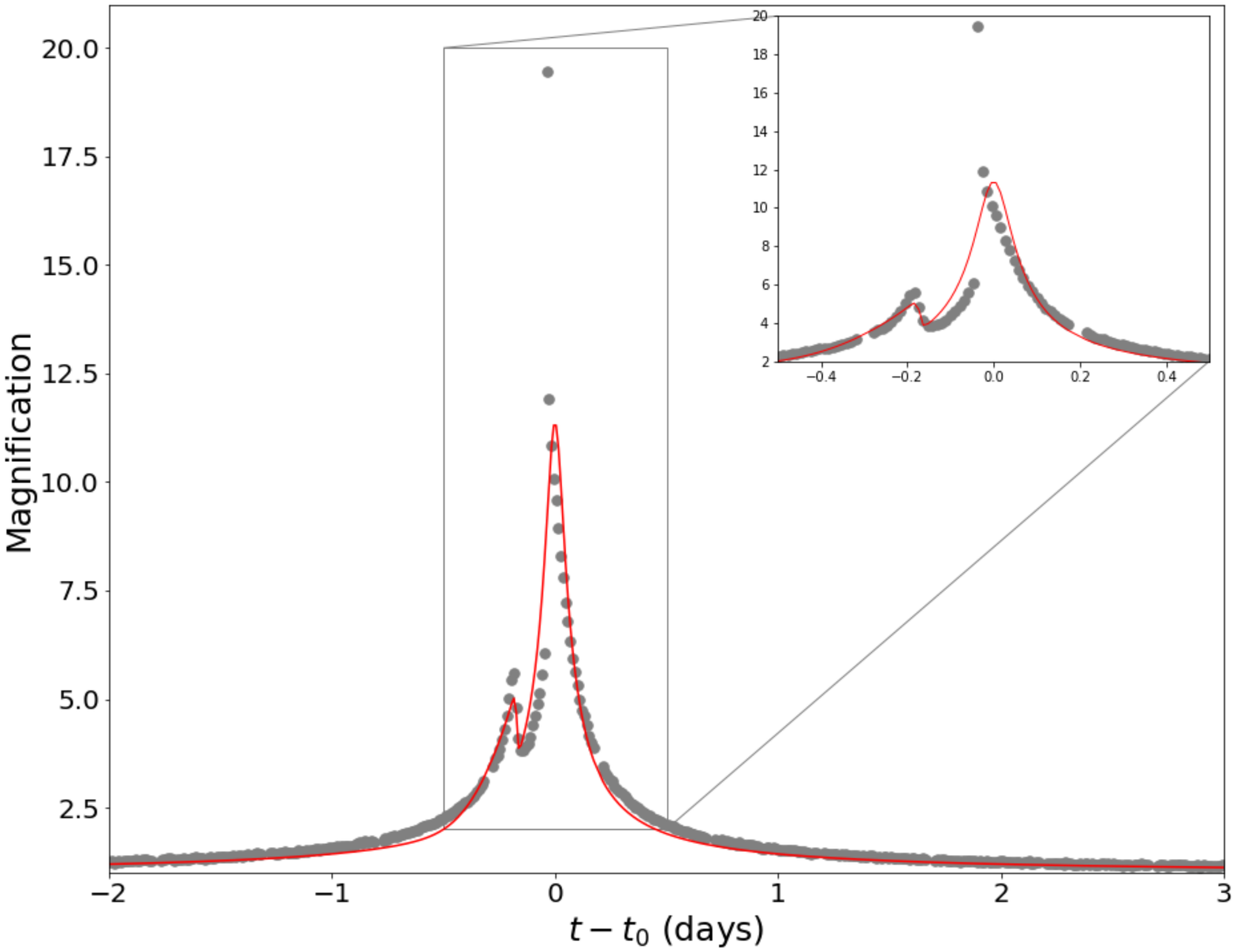}}
          \caption{An example of the PSPL plus the busy function to the light curve of a double-peaked event with $s=1.11$ and $q=0.0008$. The busy function provides a good estimate of the location and duration of the event, with the estimated system parameters of $s=1.02$ and $q=0.007$. }
          \label{fig:planetary_ex3}
        \end{figure}

Figures \ref{fig:planetary_ex1}, \ref{fig:planetary_ex2}, and \ref{fig:planetary_ex3}, show three examples of the PSPL plus the busy function fitted to planetary light curves. Figure \ref{fig:planetary_ex1} shows the fit for a system with one deviation in its PSPL residual where the true projected separation of the system is smaller than unity. The algorithm first detects sharp peaks or troughs, and after fitting the busy function to the light curve residuals from the PSPL fit, we then obtain the residual of the busy function fit. At the times where a single peak is found, we still cannot conclude that the event includes a major image perturbation. At this point, if the sum of the flux of negative points in the busy function fit residual is larger than the sum of the flux of the positive points, the system will be considered to be affected by a minor image perturbation, and $s$ will be chosen to be less than one. Otherwise, $s$ will be larger than one. The example in Figure \ref{fig:planetary_ex1}, shows an event with a sharp positive deviation that appears to be a major image perturbation; however, the preceding trough indicates that this a minor image perturbation, and the algorithm correctly decides that $s$ is smaller than one. The true parameters for this event are $s=0.63$ and $q=0.0003$ and its fitted values are $s=0.64$ and $q=0.0002$. 

Figure \ref{fig:planetary_ex2} shows the PSPL plus the busy function fitted to a double-peaked deviation with $s<1$. The busy function fit may appear to be a poor fit to the deviations but this fact is not considered a failure, and in fact helps with determining whether it is a major or a minor image perturbation. After the secondary check, the algorithm decides to correctly choose $s$ smaller than one. The true parameters for this event are $s=0.62$ and $q=0.0001$ and the fitted values are $s=0.59$ and $q=0.0003$.  Figure \ref{fig:planetary_ex3} shows another double-peaked event that has $s>1$.  The true parameters for this event are $s=1.11$ and $q=0.0008$ and the fitted values are $s=1.02$ and $q=0.007$. 

\begin{figure*}[t]
          \centerline{\includegraphics[scale=0.4]{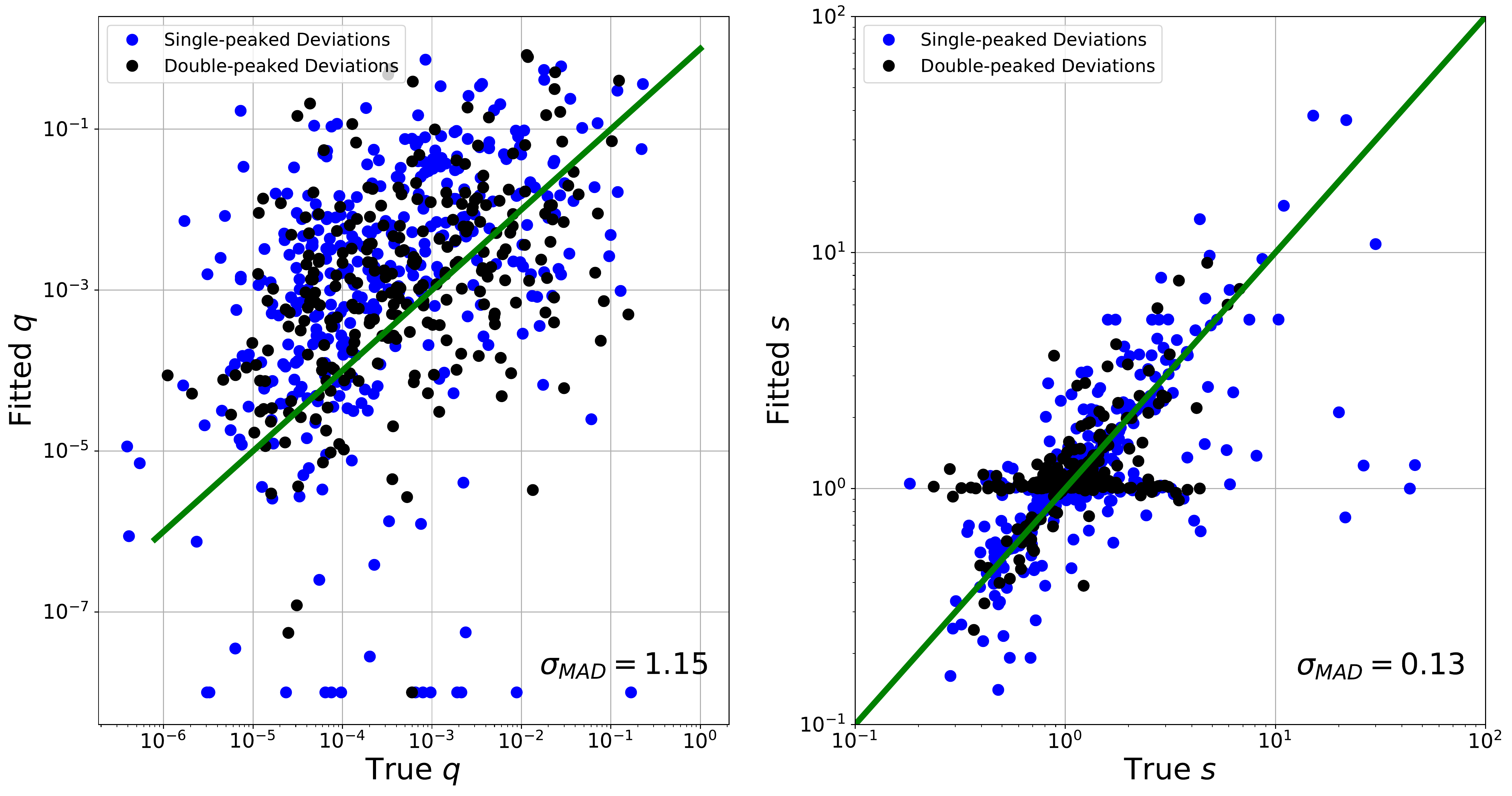}}
          \caption{Left panel shows plot of fitted values of $q$ versus the true values. Fitted values of $q$ are mostly within one order of magnitude of the true values for the PSPL plus busy function fits}. The right panel shows fitted $s$ versus the true values. Most values of $s$ are well estimated by the method. The fitted values close to one occur for events caused by the central caustics for which our formalism breaks down. Blue dots are events without caustic crossings, and black dots are events with caustic crossings.
          \label{fig:q_s_final_plot}
        \end{figure*}

In order to evaluate the performance of the algorithm, we compare the calculated values of $s$ and $q$ with their true values as demonstrated by \citet{khakpash2019fast}. Figure \ref{fig:q_s_final_plot} shows the estimated values of $s$ and $q$ plotted versus their true values. Most values of $s$ are well estimated by the method. The fitted values close to one arise from events caused by the central caustics. While the estimated Values of $q$ show a lot of scatter compared to the true value, they are generally estimated to within about one order of magnitude of the true values. Note that this algorithm is slower than the algorithm presented in \citet{khakpash2019fast}, but it fits a broader range of events.

\subsection{Chebyshev Polynomials Fit}\label{subsec:cheby}

\citet{di1997identifying} suggests that unlike binary-lens microlensing light curves that are significantly perturbed, smooth binary-lens events can be easily misclassified, and therefore, it is useful to develop methods that can distinguish between this type of microlensing events and other types of similar variabilities. The method should work fast and be effective at marking light curves with smooth features. One of the approaches to approximate a function is to expand it in form of a series of polynomials. For this purpose, \citet{di1997identifying} suggests to use the Chebyshev approximation on microlensing light curves. The idea is that features of Chebyshev polynomials are useful for capturing the smooth features of binary-lens microlensing light curves without caustic crossings. 
     
We have implemented the work of \citet{di1997identifying}, and applied it to our simulated data set. For this purpose, we first detect the peaks in the light curve, then for each peak, we choose the interval around the peak that includes the wings of the event up to a magnification of 1.06 that corresponds to an impact parameter of $2R_E$ \citep{di1997identifying}.
     
We then find the coefficients of the Chebyshev polynomials using the formulation described by \citet{press1992numerical} and expand the selected parts of the light curves by the expression in Equation \ref{eq:chebyshev}. The polynomial $T_k$ has $k+1$ number of extrema in the interval $[-1,1]$ where the Chebyshev polynomials are defined. To use this method, the time coordinates of the selected potion of our light curves should be within this interval, and so we convert the interval of time to be between $-1$ and $1$ before fitting to the Chebyshev polynomials.

     \begin{equation}\label{eq:chebyshev}
         f(x) \approx \left[\sum_{k=0}^{m} c_k T_{k}(x)\right] - \frac{1}{2}c_0
     \end{equation}

In this work, we choose the first 50 polynomials of this series ($m=50$) to fit Equation \ref{eq:chebyshev} to the light curves, and then using the coefficients $c_k$, we calculate $\Lambda = -{log}_{10}((\sum_{k=0}^{m} {(\frac{c_k}{c_0})}^2)-1)$ as a feature parameter that can be used to distinguish between different light curve types in our simulated data set. 
    
\begin{figure}
      \centerline{\includegraphics[width=1\linewidth, clip]{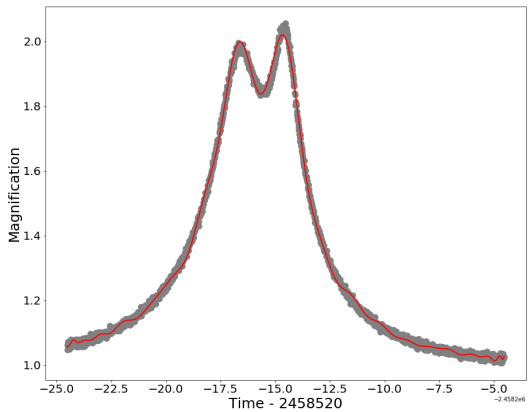}}
      \caption{The stellar binary-lens light curve is approximated with the Chebyshev polynomials of degree 50. }
      \label{fig:Chebyshevfit}
    \end{figure}

In Figure \ref{fig:Chebyshevfit}, an example of a binary-lens event approximated by the Chebyshev polynomials of degree 50 can be seen. It is important to note that the whole light curve is not approximated in this method, and only the segment containing the event excluding gaps is selected. This is because the Chebyshev approximation is good for approximating functions that have finite number of extrema in the interval of $[-1,1]$, and a non-varying light curve is not usually well fitted by this approximation.

With this method, we now need to define Chebyshev parameters that are different for various types of light curves. Here, we use $\Lambda$ and the first six even coefficients as the parameters used to distinguish between different types of light curves. The reason for that is that the coefficients would decrease rapidly when going to higher order terms in the polynomials, and therefore, the first coefficients would be dominant. Also, since the microlensing events are closer to being an even function rather than an odd function, the expansion only involves the even terms in the Chebyshev polynomials. 
     
\begin{figure}
  \centerline{\includegraphics[width=1\linewidth, clip]{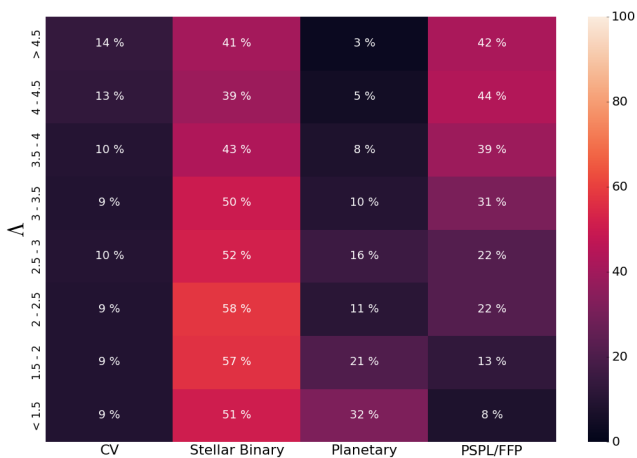}}
  \caption{ This figure shows the breakdown by variability type for each range of values of the Chebyshev feature $\Lambda$. $\Lambda$ is equal to $-{log}_{10}((\sum_{k=0}^{m} {(\frac{c_k}{c_0})}^2)-1)$, where all coefficients are normalized by the first coefficient $c_0$.}
  \label{fig:Chebyshev_ck2}
\end{figure}

We then check how well different values of $\Lambda$ match to different light curve types. Figure \ref{fig:Chebyshev_ck2} shows that the value of $\Lambda$ correlate fairly well with four different light curve types. Note that the fractions are calculated for a given range of $\Lambda$, across each light curve types, horizontally in the figure. Although the fractions of different variability types in the simulated Roman data set are not equal, there are large numbers of each variability type, so the distinctions seen in Figure \ref{fig:Chebyshev_ck2} are still quite robust. We therefore expect that the continuous parameter $\Lambda$, can be a useful input for a ML algorithm (such as a Random Forest) to classify event types.
     
In addition to the feature $\Lambda$, we also follow the suggestion of \cite{di1997identifying}, to include the individual even-numbered coefficients of the Chebyshev polynomial as classification features.

\section{An Algorithmic Approach}\label{sec:Approach}

In this section, we introduce a suggested algorithmic approach to use the produced features as input to ML classifiers. The goal of this section is to provide an efficient procedure to use these features to train classifiers on a high-cadence data set and detect different types of microlensing light curves. In Table \ref{table:Summary}, we have a summary of the tools in our package and the associated features. Each of these tools can be either applied to the whole dataset or a subset of it, and there may be more than a single algorithmic approach to use these features.

\begin{table*}
\caption{Summary of the introduced algorithms and their resulting features.}
\centering
\begin{tabular}{|l|l|}
  \hline
     {\bf Algorithms} & {\bf Produced Features}\\ \hline\hline
     \hyperref[subsec:PeakFinder]{Peak Finder} & Number of the Peaks\\ \hline
     \hyperref[subsec:gpspl]{G-PSPL Fit} & ${\chi}^2$: Goodness of the G-PSPL fit\\ \hline
     \hyperref[subsec:symcheck]{Symmetry Check} & $\beta$: A measure of asymmetry of the peak\\ \hline
     \hyperref[subsec:trap]{Trapezoidal Function Fit} & $\kappa$: Ratio of the duration of the flat top of the trapezoidal function to total duration\\ 
      & $t_{E,trap}$: Duration of the trapezoidal function\\ \hline
     \hyperref[subsec:cauchy]{Cauchy Distribution Fit} & $\psi$: The difference between the goodness of PSPL and Cauchy fits\\
      & $b$: A measure of the flatness of the top\\ \hline
     \hyperref[subsec:busy]{Planetary Parameter Finder} & $s$: Projected mass ratio in units of Einstein radius\\
      & $q$: Mass ratio\\ \hline
     \hyperref[subsec:cheby]{Chebyshev Polynomials Fit} & $\Lambda$: Sum of the square roots of the Chebyshev coefficients\\
      & $a_{2}$, $a_{4}$, $a_{6}$, $a_{8}$, $a_{10}$: First five even coefficients of Chebyshev polynomials \\ \hline
    
\end{tabular}  
\label{table:Summary}

\end{table*} 

In a paper under development (Khakpash et al, in preparation), we are implementing an algorithmic approach to employing the features discussed in this paper to comprehensively search for microlensing events, classify them by type, and derive preliminary system parameters.  For any set of lightcurve features, there are numerous ways to conduct classification and our approach is by no means certain to be optimal in every way.  We encourage other astronomers to make use of the features described here to develop independent classification methods.

The particular structure or sequence of a classification approach can vary. One might use all the features to detect all the categories at once, or alternatively, find particular classes by doing a step-by-step classification. In an ideal case where there are thousands examples of each class across a full range of empirical properties of each feature, it is likely better to use all the features to find all of the classes in a one-step classification.  Since in this case we have very limited examples of some of the classes, we recommend a step-by-step classification approach.

The first step of classification is to distinguish microlensing light curves from other types of variability (e.g., CV in our dataset). We refer to this step as classification step I. Although our goal in this work is not focused on this set, we believe some of the features identified in this work can be used to improve the current existing classifiers focused on this task. We have tested this type of classification using the features such as number of the peaks found by the Peak Finder, $\Lambda$, $a_{2}$, $a_{4}$, $a_{6}$ produced by the Chebyshev fit, and ${{\chi}^2}_{PSPL}$, ${ t_{E,PSPL} }$, $\beta$ generated by the G-PSPL fit and the Symmetry Check.

Assuming we have identified all the microlensing light curves, the next type of classification is to classify them into single-lens versus binary-lens events which we refer to as step { \it II}. Note that in a real dataset, multi-lens events also exist which can either be added as a category or can be included in one category along with the binary-lens events. We use the same feature as in step I excluding the number of peaks. 

Once we have single-lens and binary-lens events, we classify the binary-lens events into stellar binary-lens and planetary binary-lens systems (classification step III) using ${{\chi}^2}_{PSPL}$, ${ t_{E,PSPL} }$ produced by the PSPL fit and $s$, and $q$ produced by the PSPL plus busy fit.

Furthermore, single-lens events can be classified into classes of isolated short timescale events likely caused by free-floating planets or planets on very wide orbits, stellar PSPL, and stellar FSPL events (Classification step {\it IV}). We suggest using $\kappa$, $\Delta \tau_{full}$, $\psi$, and $b$ produced by the Trapezoidal Fit and the Cauchy/PSPL Fit to obtain better results. This step will be investigated in the future work.

\begin{figure*}[!htbp]
	\centering
	\includegraphics[width=0.8\linewidth, clip]{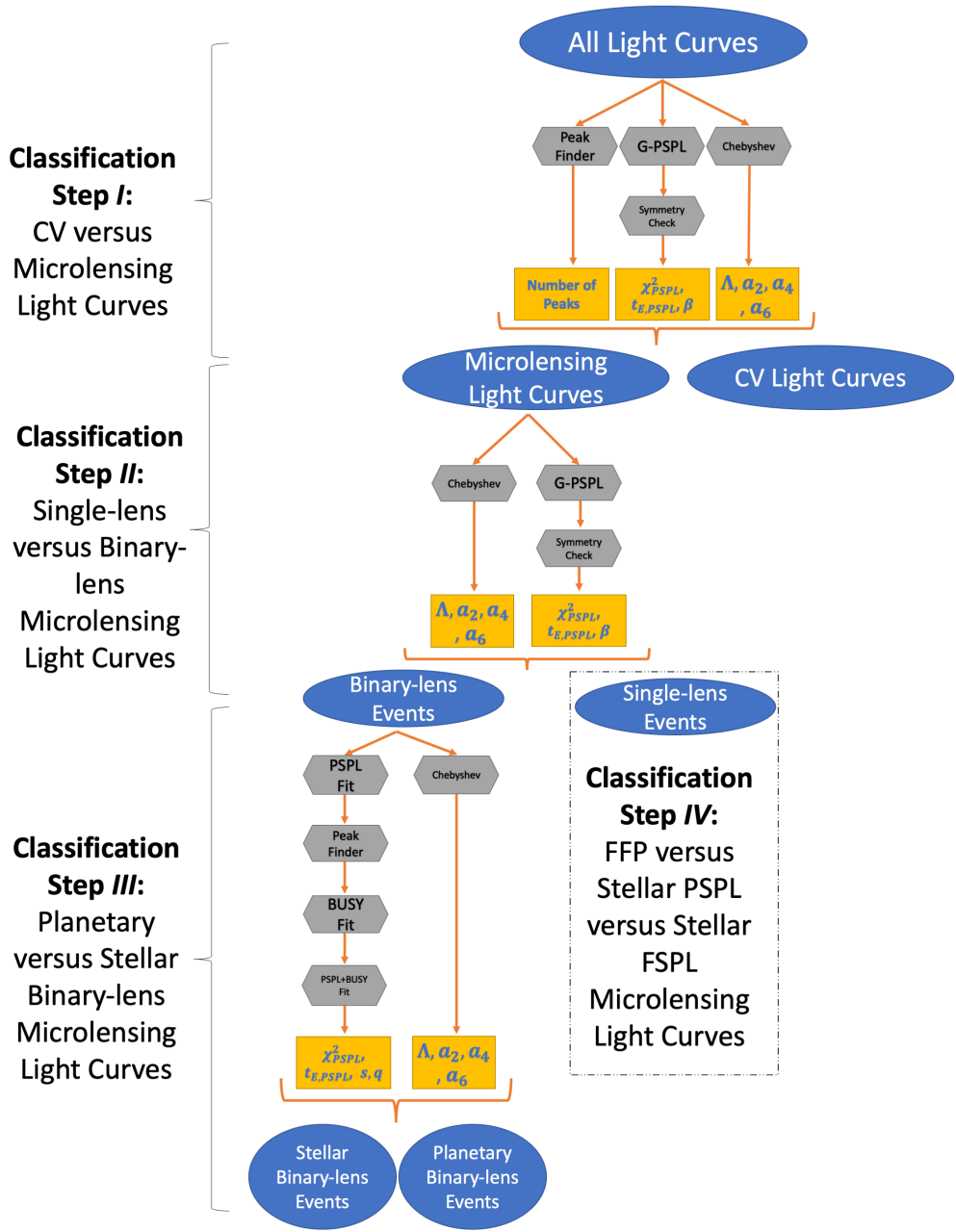}
	\caption{ This diagram shows an algorithmic approach for using the features described in this paper} as input to ML classifiers.
	\label{fig:diagram}
\end{figure*}

Figure \ref{fig:diagram} displays a diagram showing our suggested algorithmic approach to use the different features produced by our package. We have selected this approach to optimize the classification results considering our small dataset. We have tested classification steps I, II, and III with four ML classifiers including a {\it k}-nearest neighbors classifier, a decision tree classifier, a random forest classifier and a neural network classifier. The preliminary results for that are presented in the next section.

\section{Preliminary Testing of ML Algorithms}\label{sec:ML_test}

As stated above, there are many ways to implement an approach to identify and characterize microlensing events with classification based on light curve features.  We are developing a thorough investigation into that question (Khakpash, et al. in preparation), but for now we present a preliminary analysis using a few simple analysis steps based on the classification approach described in \S \ref{sec:Approach}. We present the results of training four ML classifiers including a {\it k}-nearest neighbor classifier (KNN), a decision tree classifier (DT), a random forest classifier (RF), and a neural network classifier (NN) using the features we introduced in this paper. In order to test these algorithms, we follow the step-by-step scheme of Figure \ref{fig:diagram}. It is important to note again that the step {\it IV} of the classification in Figure \ref{fig:diagram} is not tested here and will be pursued in the future. At each step, we set the size of our test set to be 20\% of the whole dataset. The test set is then randomly chosen in a 5-fold cross-validation process, and the average scores are reported at the end.

In order to compare results if these classifiers, we show confusion matrices made with the test set at each step along with their Receiver operating characteristic (ROC) curves. As mentioned in Section \ref{sec:Approach}, we understand that the one-step classification is a more ideal approach, and we tested this approach with our current data set. However, the limited number of object instances in each class was insufficient for achieving an overall acceptable accuracy. Nevertheless, the isolated confusion matrices of the step-by-step classifications are valuable to evaluate the utility of the features presented in this paper. We are planning to thoroughly investigate the one-step classification in a future paper. 

At each step, we use the same dataset and features for all of the four classifiers, but we optimize their hyperparameters separately. The RF, KNN, and DT classifiers are implemented using the scikit-learn package in python \citep{pedregosa2011scikit}. The NN classifier is implemented using the Tensorflow and Keras packages in python \citep{tensorflow2015-whitepaper, chollet2015keras}. A summary of the features and hyperparameters of each classifier at each step is given in Table \ref{table:classifiers_info}. 

\subsection{Classification Step I} \label{sec:class1}
 
As shown in Figure \ref{fig:diagram}, the first step of the algorithmic approach is to detect microlensing light curves among other stellar variability. We should first note that our current light curve data set is not completely representative of what we expect from the Roman mission, since the non-microlensing type of variability in the light curves in our current dataset only includes cataclysmic variables. Most (but not all) forms of non-microlensing variability are expected to be periodic or quasi-periodic, and thus simply including CVs is a good starting point. Our dataset for this set contains 4181 light curves among which there are 3752 microlensing light curves (labeled as 1) and 429 non-microlensing light curves (labeled as 0). 

\begin{figure*}[t]
    \centering
    \subfloat[Random Forest Classifier]{%
      \includegraphics[width=0.4\textwidth]{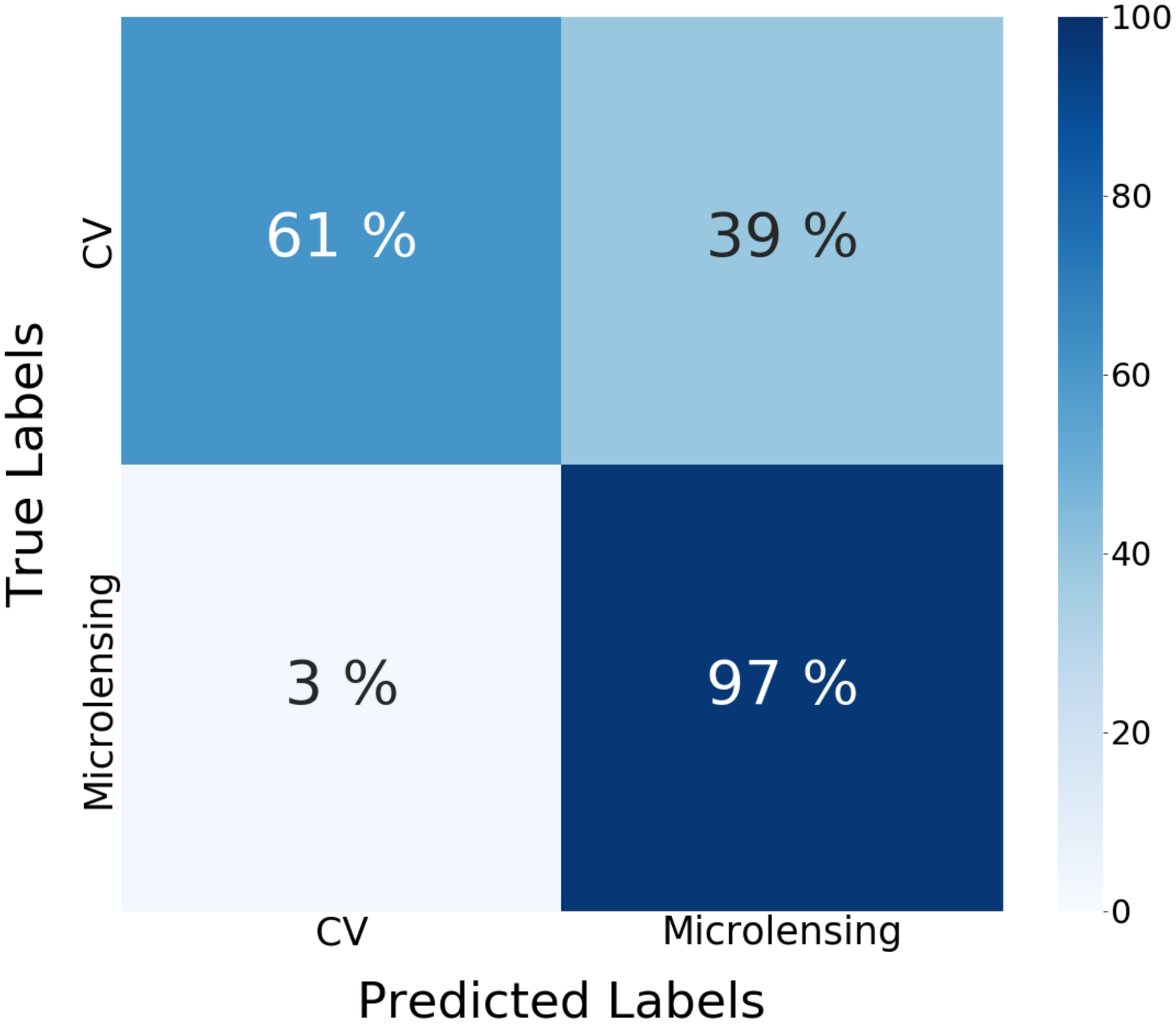}%
      \label{plot:set1_rf}%
    }\qquad
    \subfloat[Neural Networks Classifier]{%
      \includegraphics[width=0.4\textwidth]{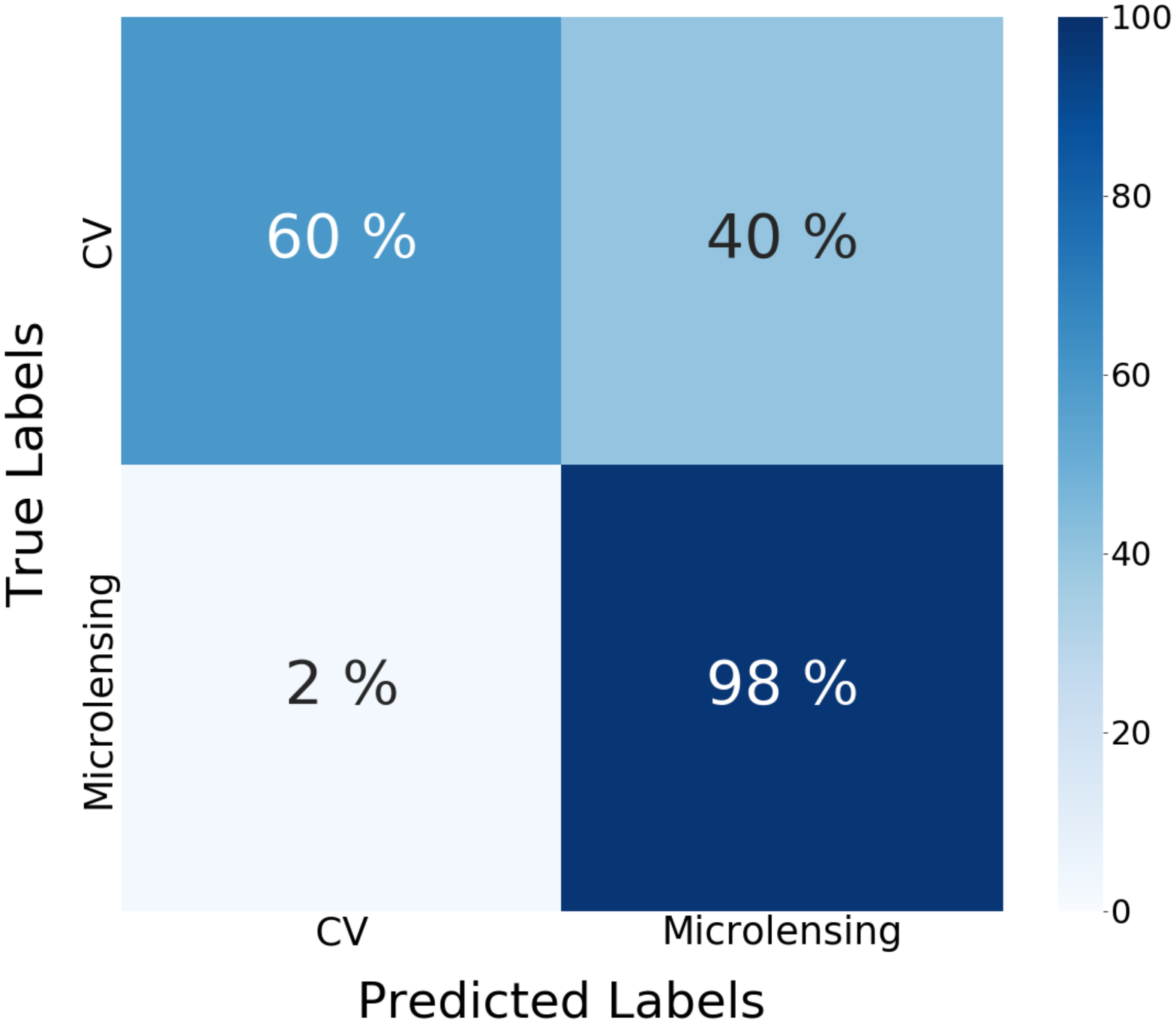}%
      \label{plot:set1_nn}%
    }\qquad
    \subfloat[Decision Tree Classifier]{%
      \includegraphics[width=0.4\textwidth]{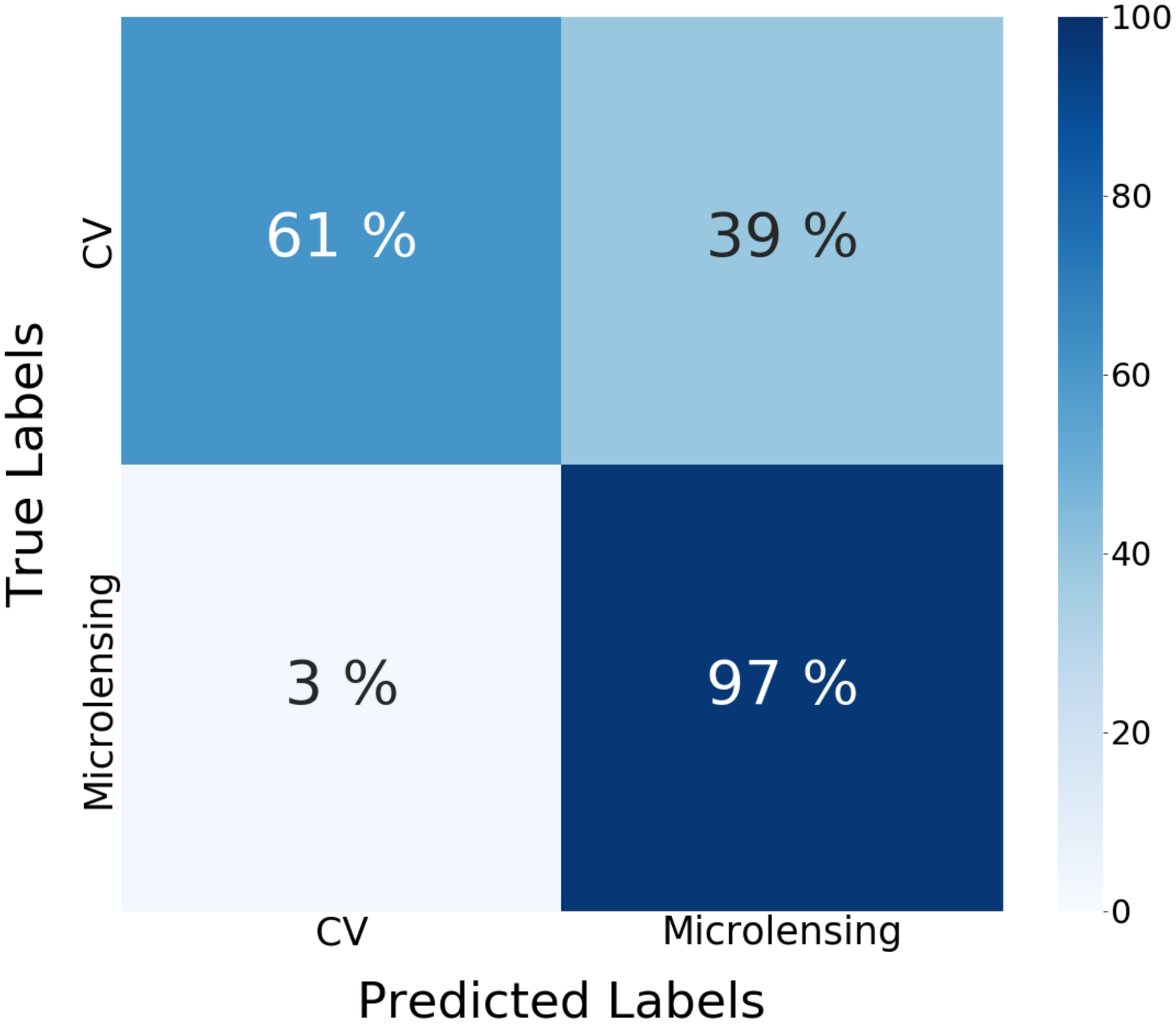}%
      \label{plot:set1_dtre}%
    }\qquad
    \subfloat[{\it K}-nearest Neighbors]{%
      \includegraphics[width=0.4\textwidth]{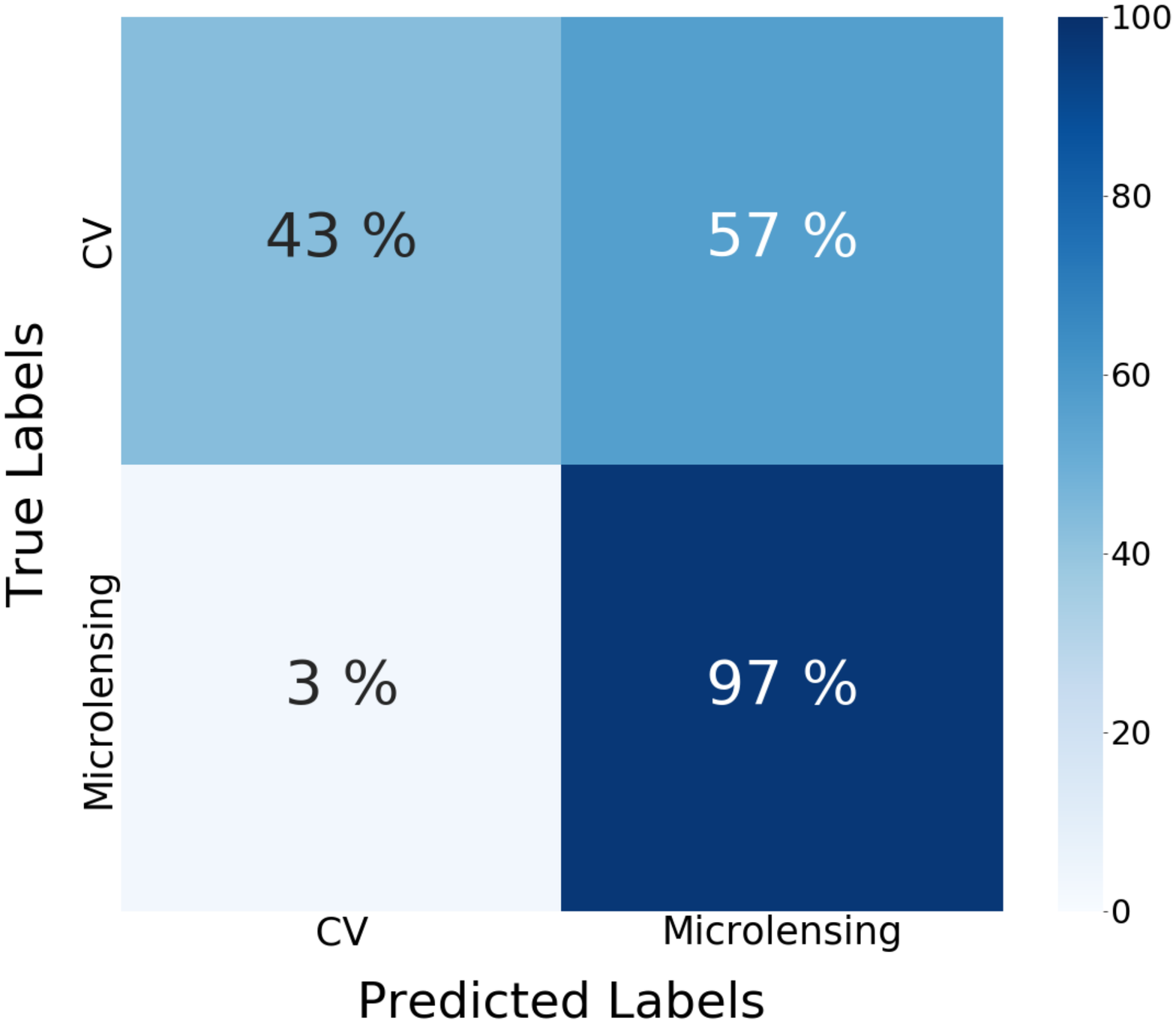}%
      \label{plot:set1_knn}%
    }
    \caption
    {\small Confusion matrices of the four trained classifiers of classification Step I. In this step, we aim at classifying all of the light curves into two classes of CV and microlensing.} 
    \label{fig:CM_set1}
    
    \end{figure*}

We find that the test set and training set accuracy for all of the four classifiers in this step are very close. A common way to evaluate the results of a classification model is to plot a confusion matrix for it. A confusion matrix is a table containing the percentages of both correctly and incorrectly classified objects for each class in the dataset. According to the confusion matrices shown in Figure \ref{fig:CM_set1}, most of ML tools can find the microlensing light curves with very small classification error, whereas, about 40\% of the CV light curves are misclassified. This could be a result of having a small training set, or incomplete hyperparameter tuning, or might be an indication of the need to include more features.

\begin{figure}[t]
    \centerline{\includegraphics[width=1.2\linewidth, clip]{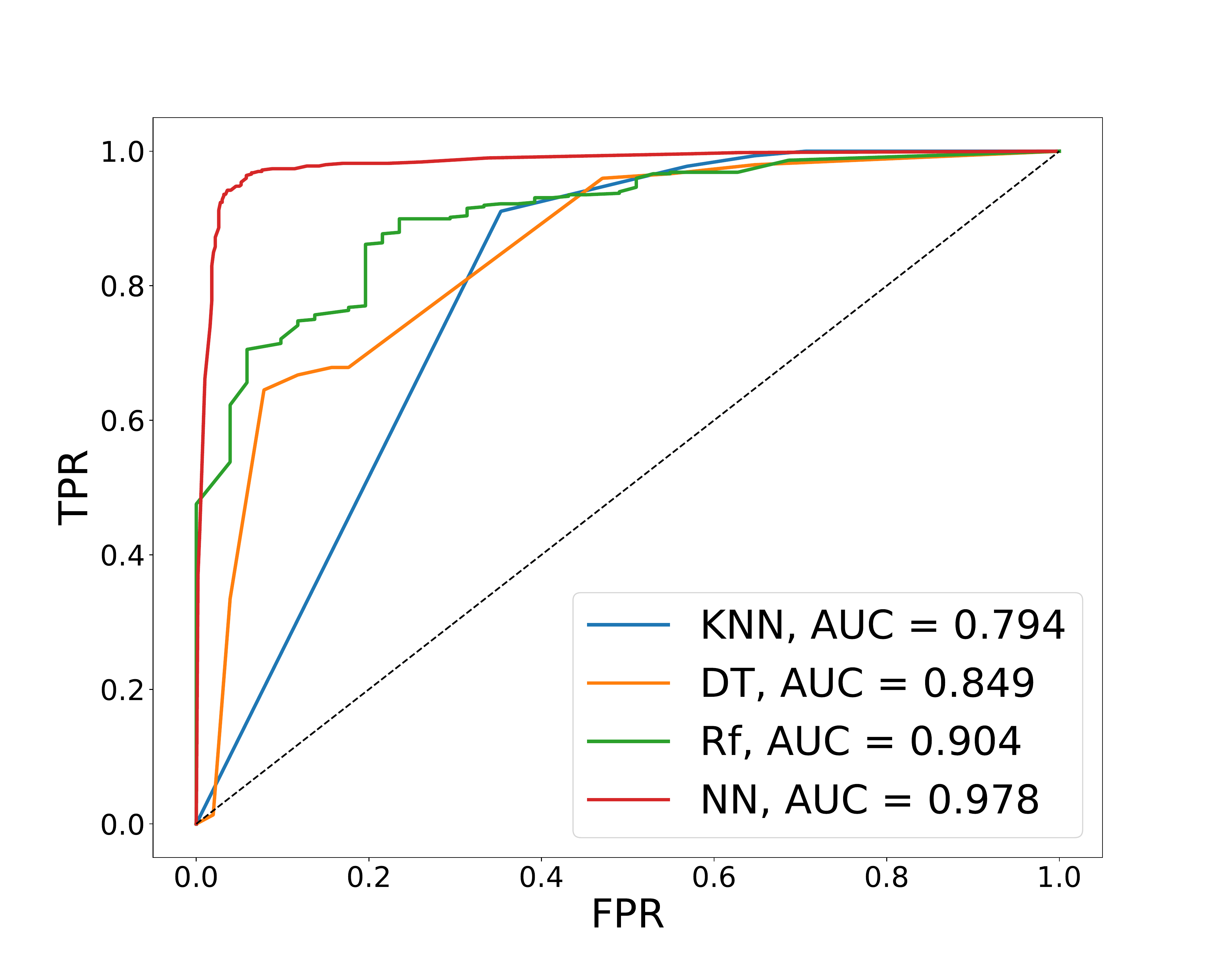}}
    \caption
    {\small ROC curves of the four trained classifiers of classification Step I along with their AUC values. In this step, we aim at classifying all of the light curves into two classes of CV and microlensing. RF and NN have the largest AUC implying that they are able to achieve higher TPR while the FPR is also low. }
    \label{fig:ROC_set1}
    
    \end{figure}

A more robust method of comparing different ML classifiers is to plot their ROC curves, and calculate their values of the Area Under the Curve (AUC). The closer the AUC is to unity, the better the performance of that classifier is. This includes plotting True Positive Rate (TPR) versus False Positive Rate (FPR) for different decision thresholds, and finding the area under that. Figure \ref{fig:ROC_set1} shows the ROC curves of the four classifiers trained in step I. The diagonal dashed line represents the ROC curve of a random classification. The ROC curves show that the NN and RF classifiers have a better performance and can achieve a higher TPR without lowering the FPR.

\subsection{Classification Step II} \label{sec:class2}

The second classification step is distinguishing between single-lens and binary-lens microlensing light curves as shown in Figure \ref{fig:diagram}. Our dataset for this step contains 4181 light curves among which there are 2143 binary-lens microlensing light curves (labeled as 1) and 1626 single-lens microlensing light curves (labeled as 0). For this step, we find that DT and RF have similar training and test accuracy and seem to work better than NN and KNN, although NN seems to work much better than KNN. 

\begin{figure*}[t]
    \centering
    \subfloat[Random Forest Classifier]{%
      \includegraphics[width=0.4\textwidth]{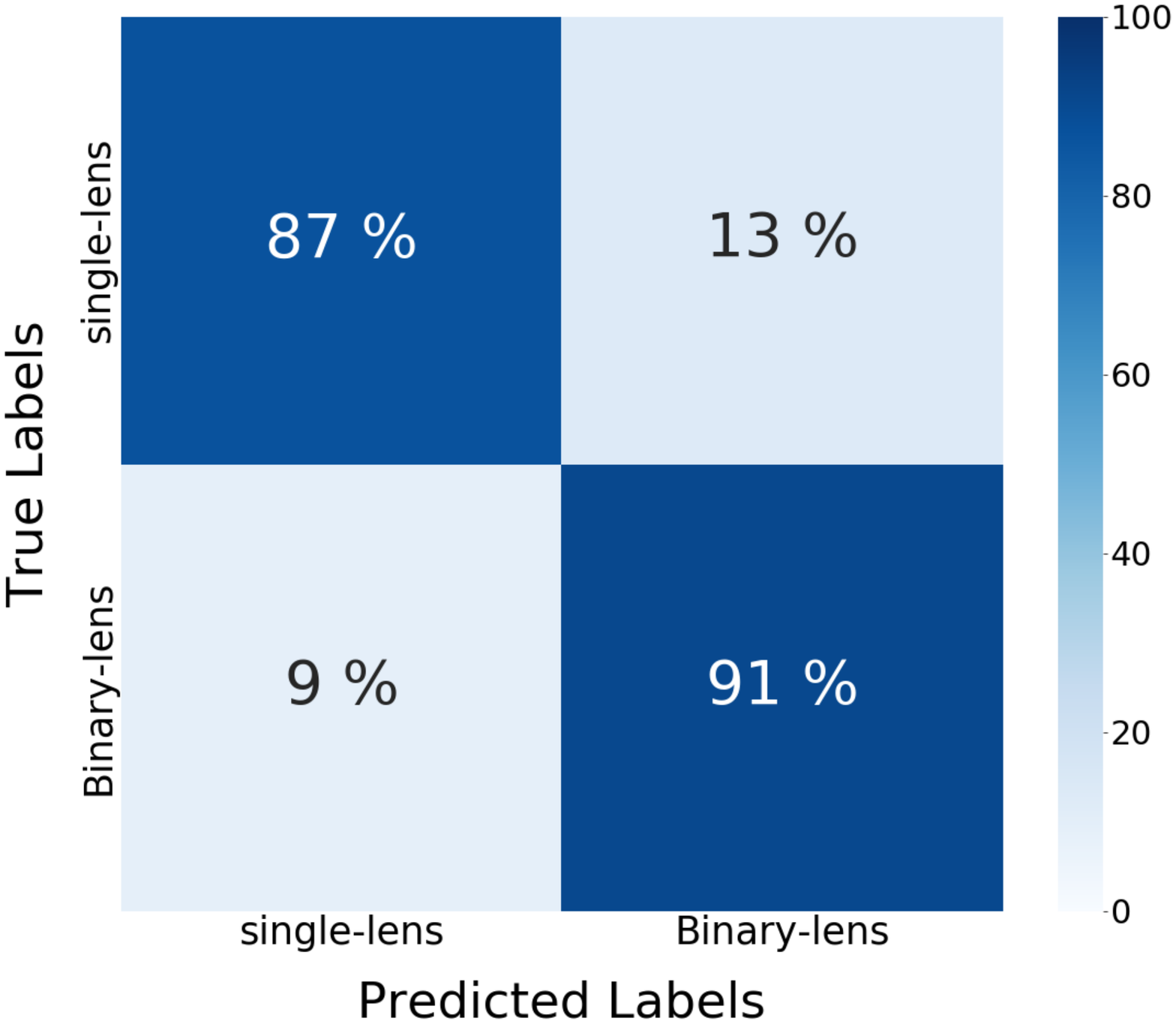}%
      \label{plot:set2_rf}%
    }\qquad
    \subfloat[Neural Networks Classifier]{%
      \includegraphics[width=0.4\textwidth]{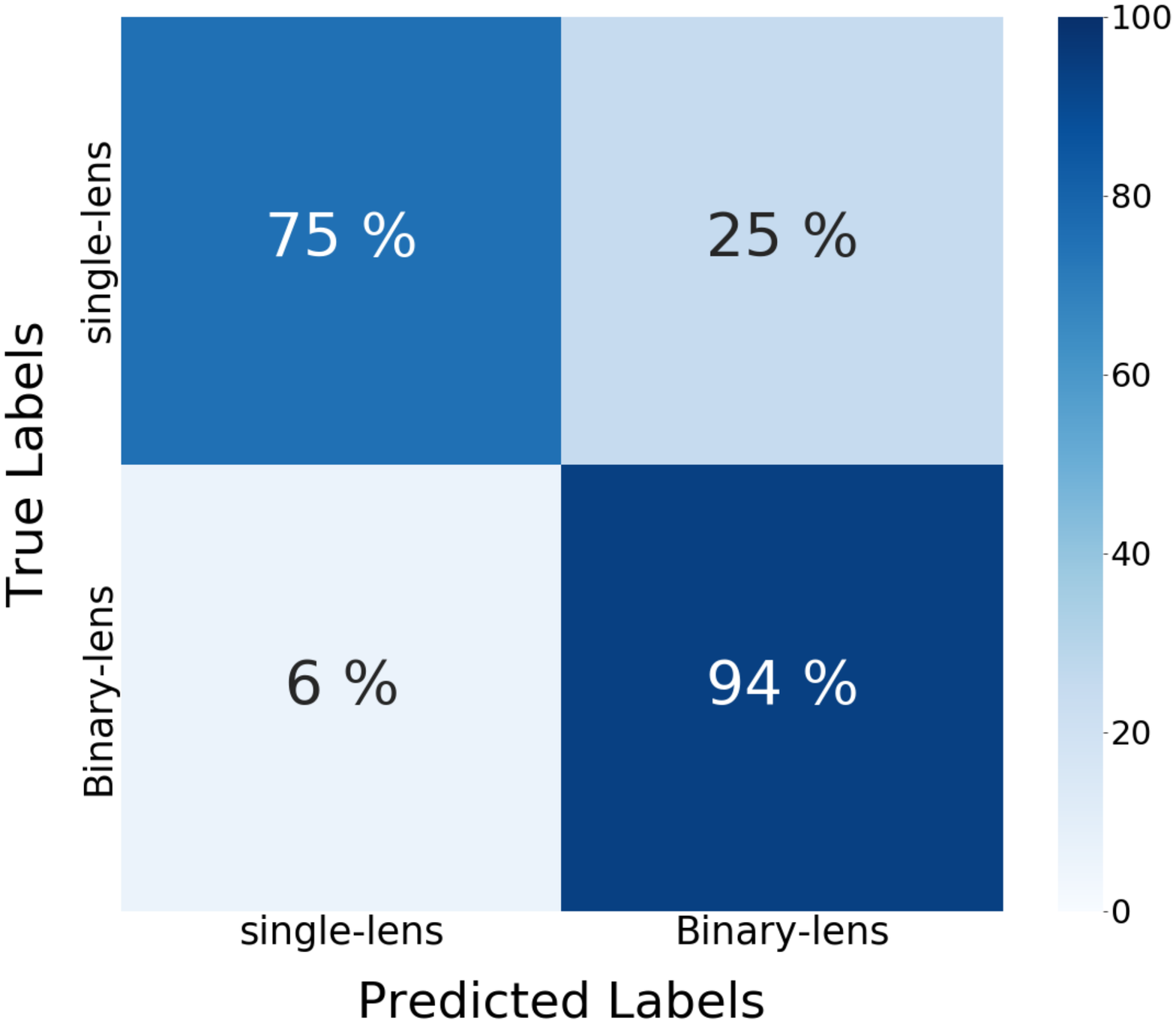}%
      \label{plot:set2_nn}%
    }\qquad
    \subfloat[Decision Tree Classifier]{%
      \includegraphics[width=0.4\textwidth]{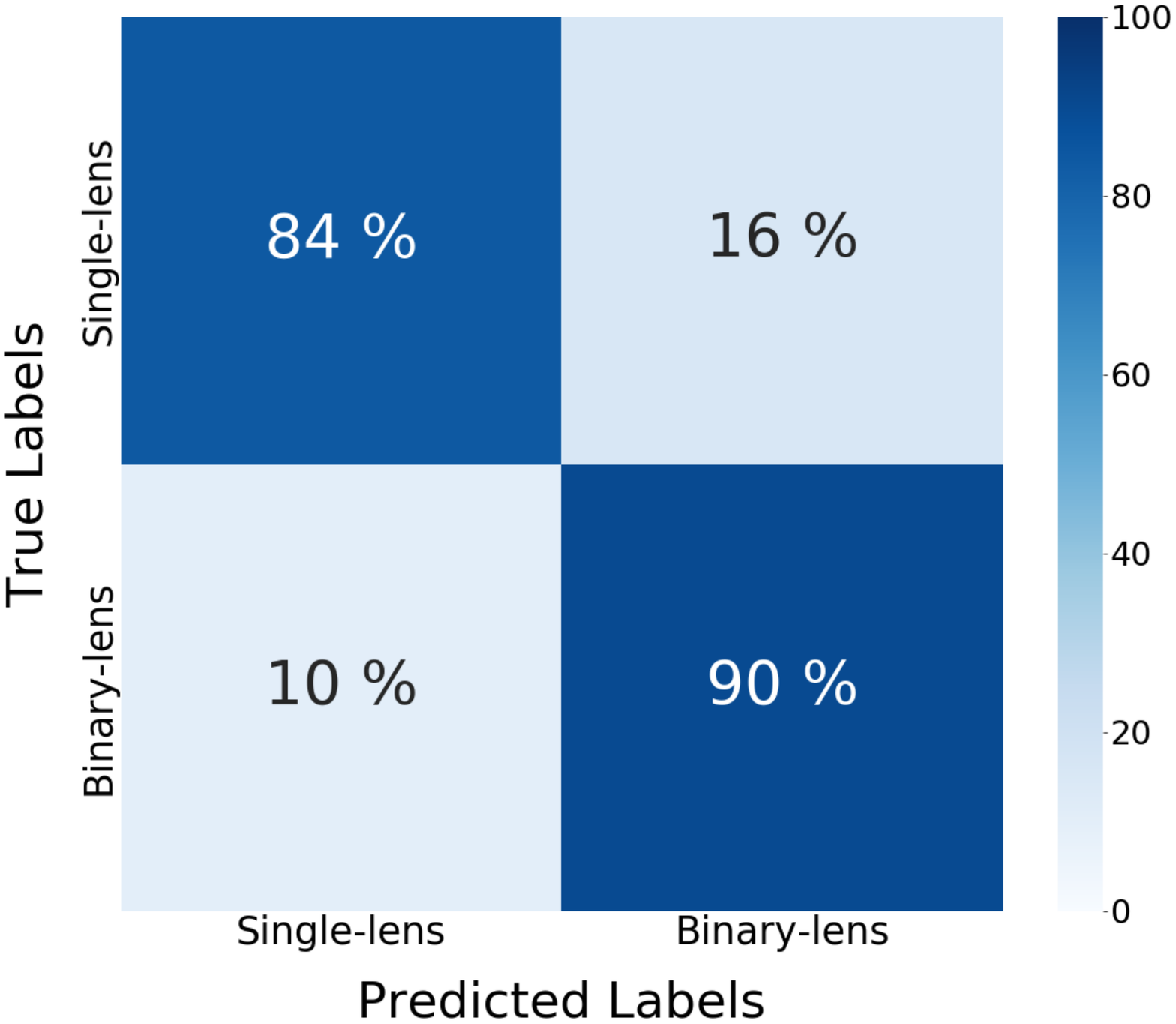}%
      \label{plot:set2_dtre}%
    }\qquad
    \subfloat[{\it K}-nearest Neighbors]{%
      \includegraphics[width=0.4\textwidth]{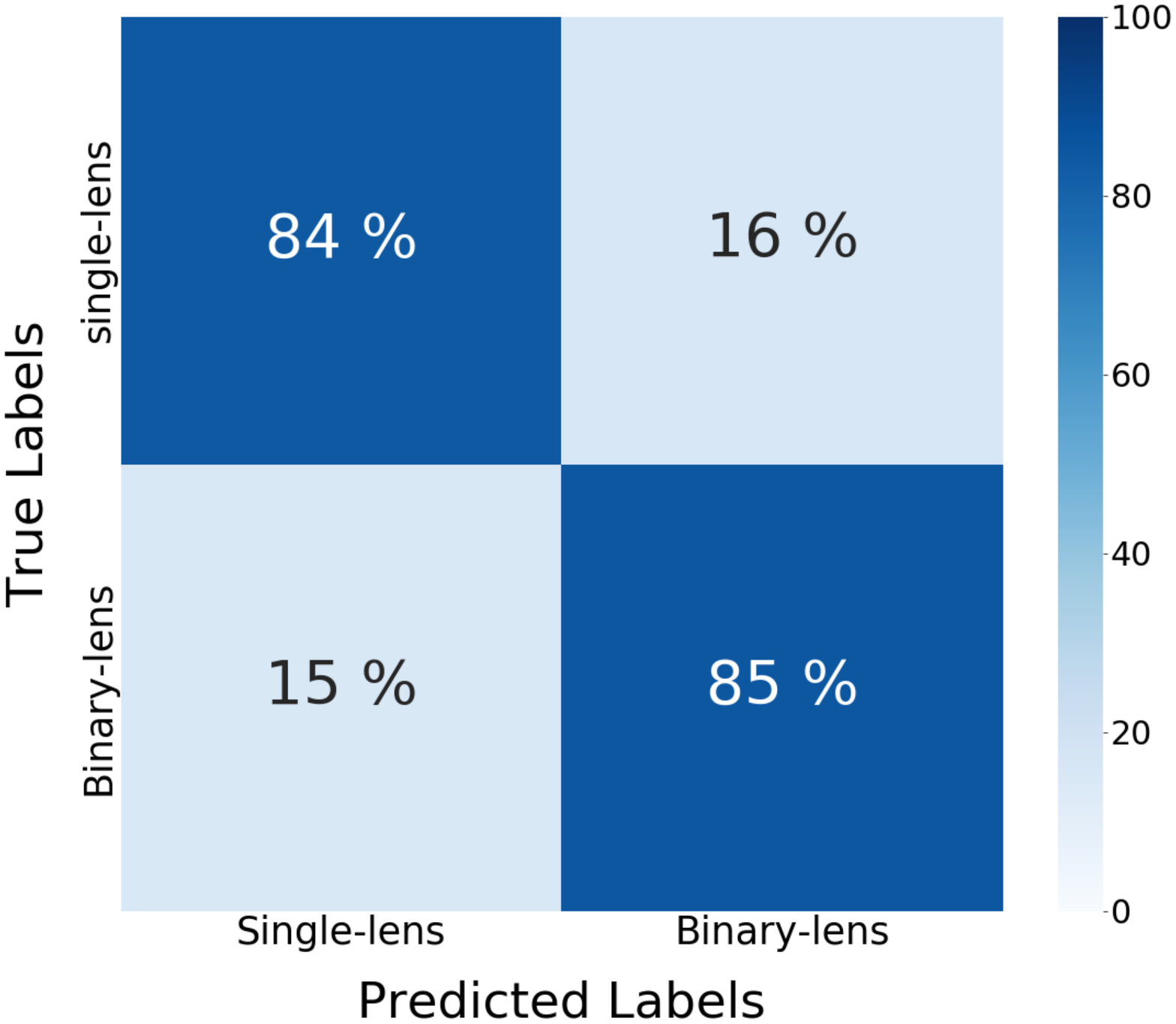}%
      \label{plot:set2_knn}%
    }
    \caption
    {\small Confusion matrices of the four trained classifiers of classification Step II. At this stage, we assume that the microlensing light curves are already detected, and our goal is to classify them into groups of single-lens and binary-lens microlensing light curves.} 
    \label{fig:CM_set2}
    \end{figure*}

\begin{figure}[t]
    \centerline{\includegraphics[width=1.2\linewidth, clip]{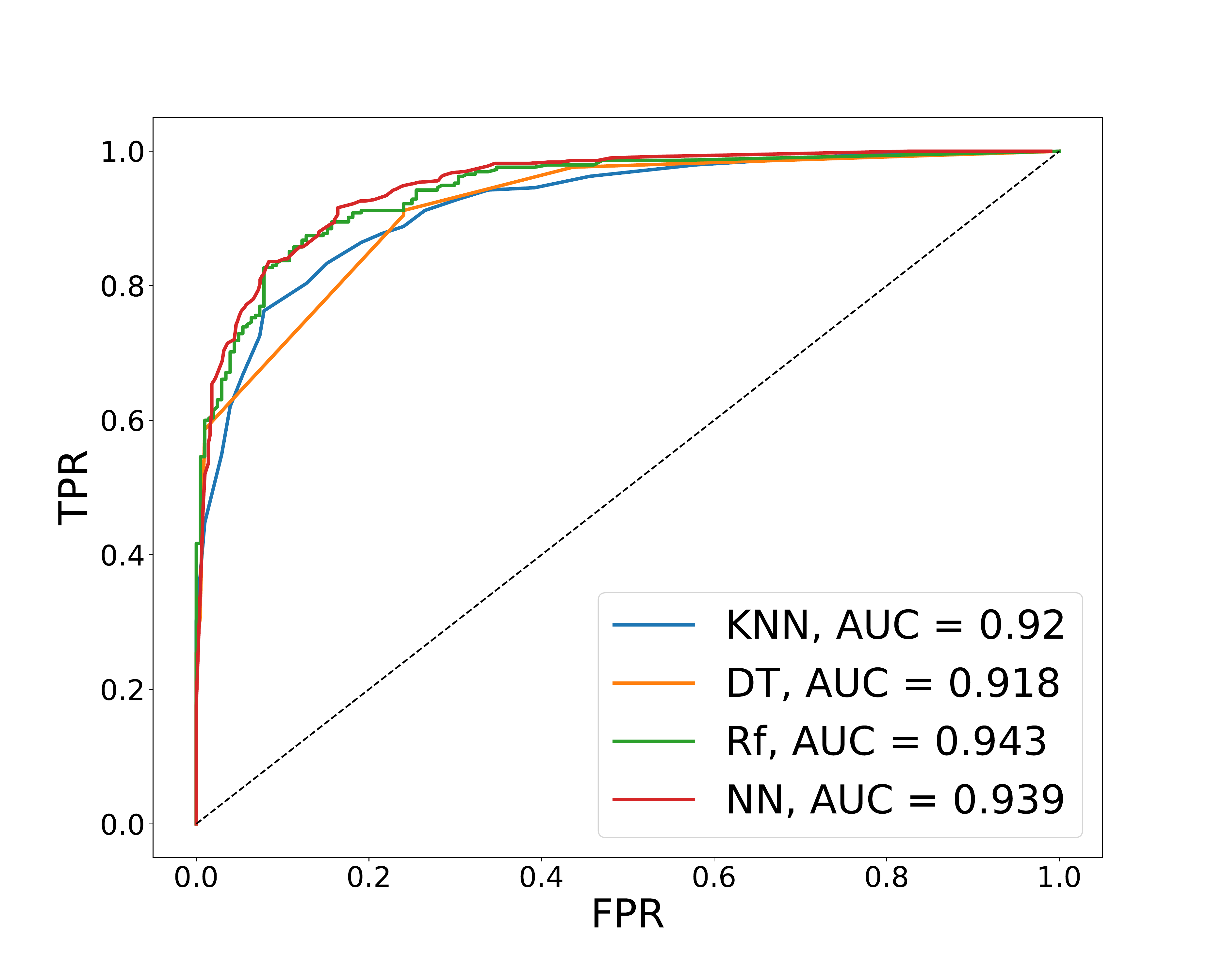}}
    \caption
    {\small ROC curves of the four trained classifiers of classification Step II along with their AUC values. In this step, we are classifying the microlensing light curves into groups of single-lens and binary-lens microlensing light curves. The dataset is more balanced in this step and all of the classifiers seem to work well. RF and NN still show a better performance.} 
    \label{fig:ROC_set2}
    
    \end{figure}

Figure \ref{fig:CM_set2} shows confusion matrices of the four classifiers trained for this step. The confusion matrices suggest that RF works better at predicting both labels, whereas NN works best at finding a larger fraction of the binary-lens light curves. Figure \ref{fig:ROC_set2} shows the ROC curves and AUC values of this step. The classifiers are overall working better in this step mainly because the dataset is more balanced here.

\subsection{Classification Step III} \label{sec:class3}

After finding the binary lens light curves, the next classification step is distinguishing between stellar binary-lens and planetary binary-lens microlensing light curves as shown in Figure \ref{fig:diagram}. Our dataset for this step contains 2074 light curves among which there are 688 planetary binary-lens microlensing light curves (labeled as 1) and 1386 stellar binary-lens microlensing light curves (labeled as 0).

This classification step is particularly important in this context since planetary binary-lens systems are the ones that astronomers would like to distinguish from the rest of the dataset. In our tested example the number of light curve is lower than the previous classification steps and this decreases the accuracy of the classifiers. Because of this, we find that the overall accuracy values of this step are smaller than the previous steps. Additionally, stellar and planetary binary-lens systems are much less distinguishable from each other compared to the previous tasks, and for this reason the simpler algorithms of DT and KNN appear to have lower accuracy. RF and NN have higher overall accuracy, but NN has a higher test accuracy which results in a larger fraction of the test set being correctly labeled. It seems that an algorithm like NN is more capable of distinguishing between these two categories which is expected as NN is theoretically more complex and is designed to find complicated patterns in a data set.

\begin{figure*}[t]
    \centering
    \subfloat[Random Forest Classifier]{%
      \includegraphics[width=0.4\textwidth]{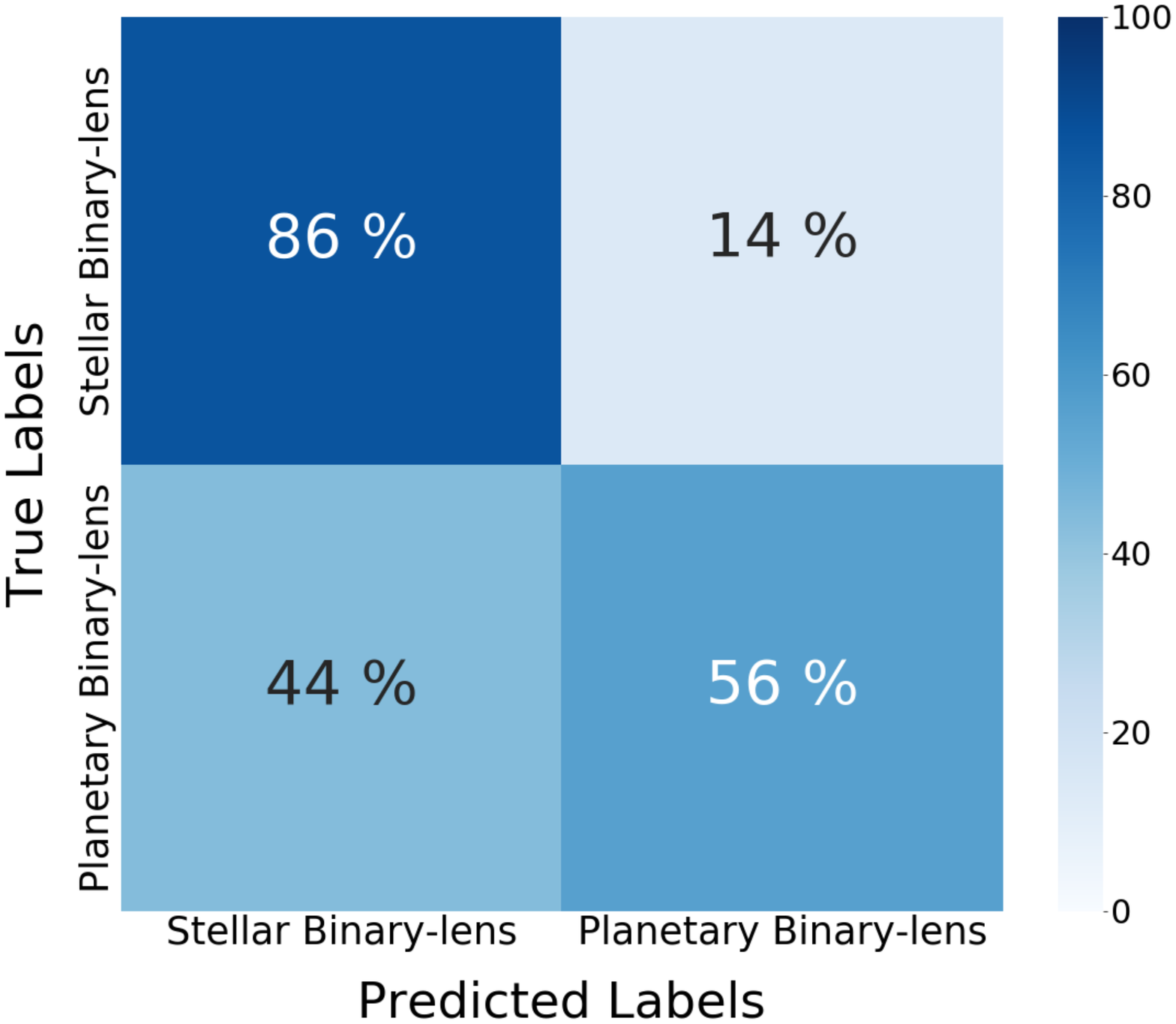}%
      \label{plot:set3_rf}%
    }\qquad
    \subfloat[Neural Networks Classifier]{%
      \includegraphics[width=0.4\textwidth]{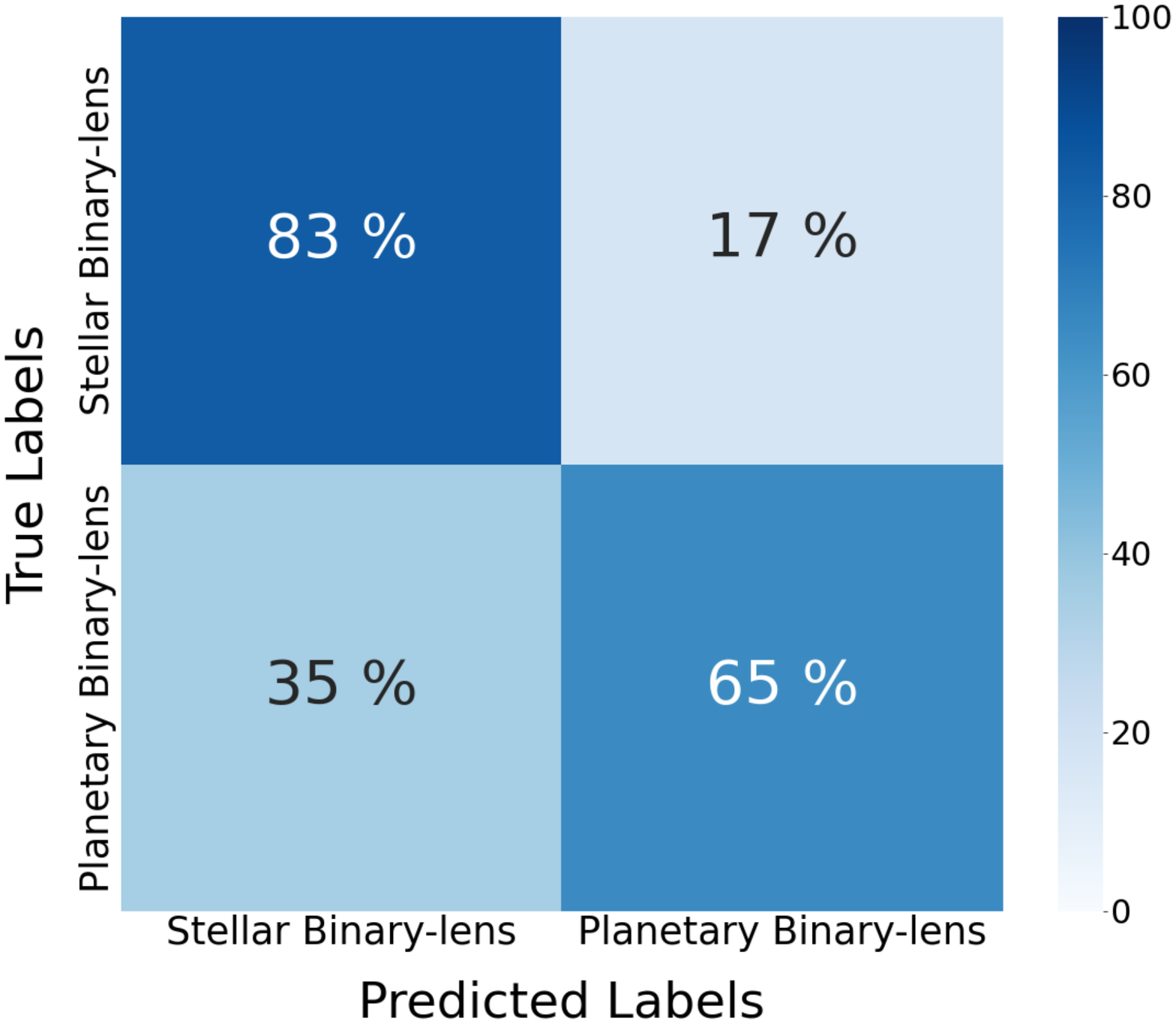}%
      \label{plot:set3_nn}%
    }\qquad
    \subfloat[Decision Tree Classifier]{%
      \includegraphics[width=0.4\textwidth]{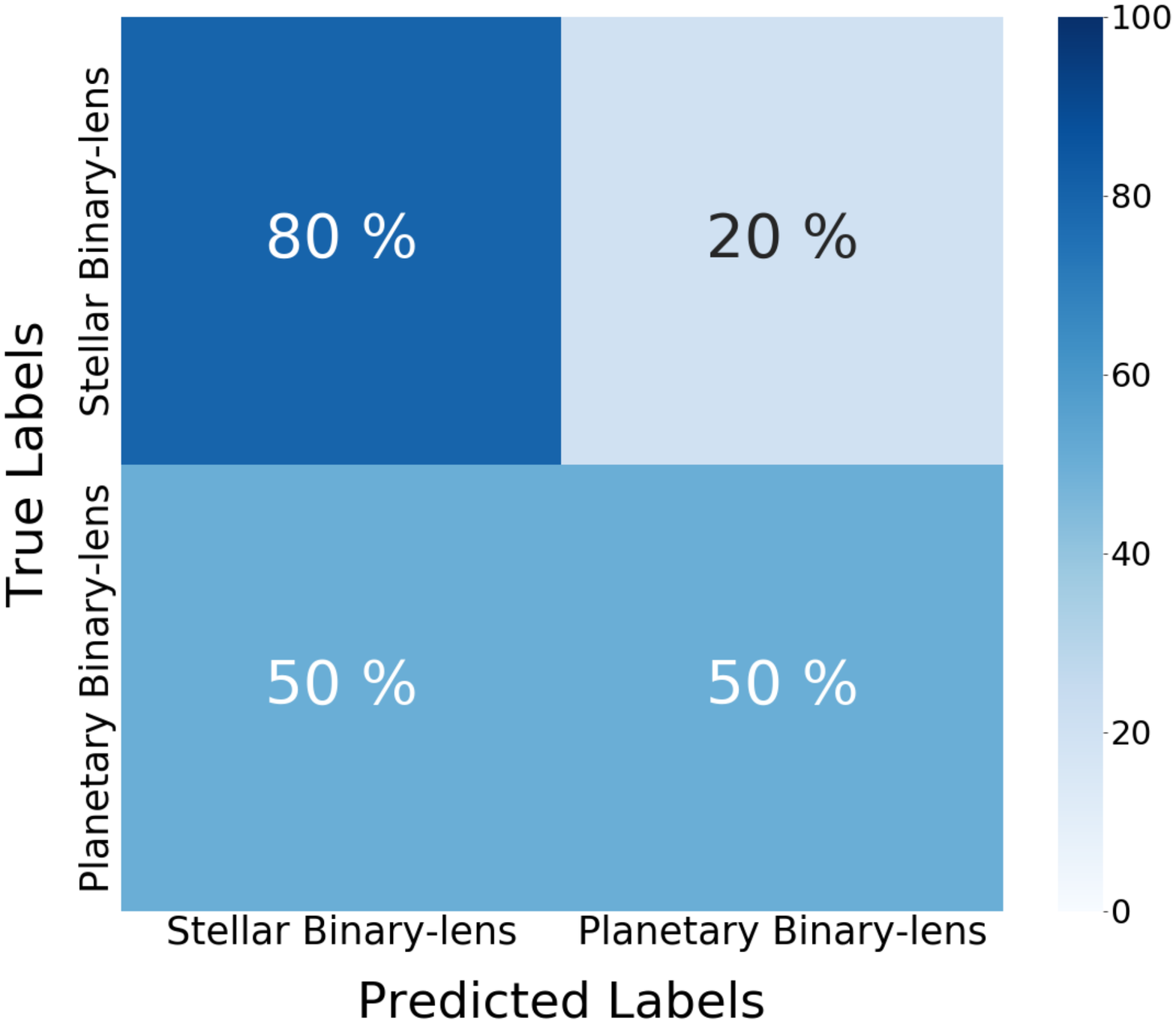}%
      \label{plot:set3_dtre}%
    }\qquad
    \subfloat[{\it K}-nearest Neighbors]{%
      \includegraphics[width=0.4\textwidth]{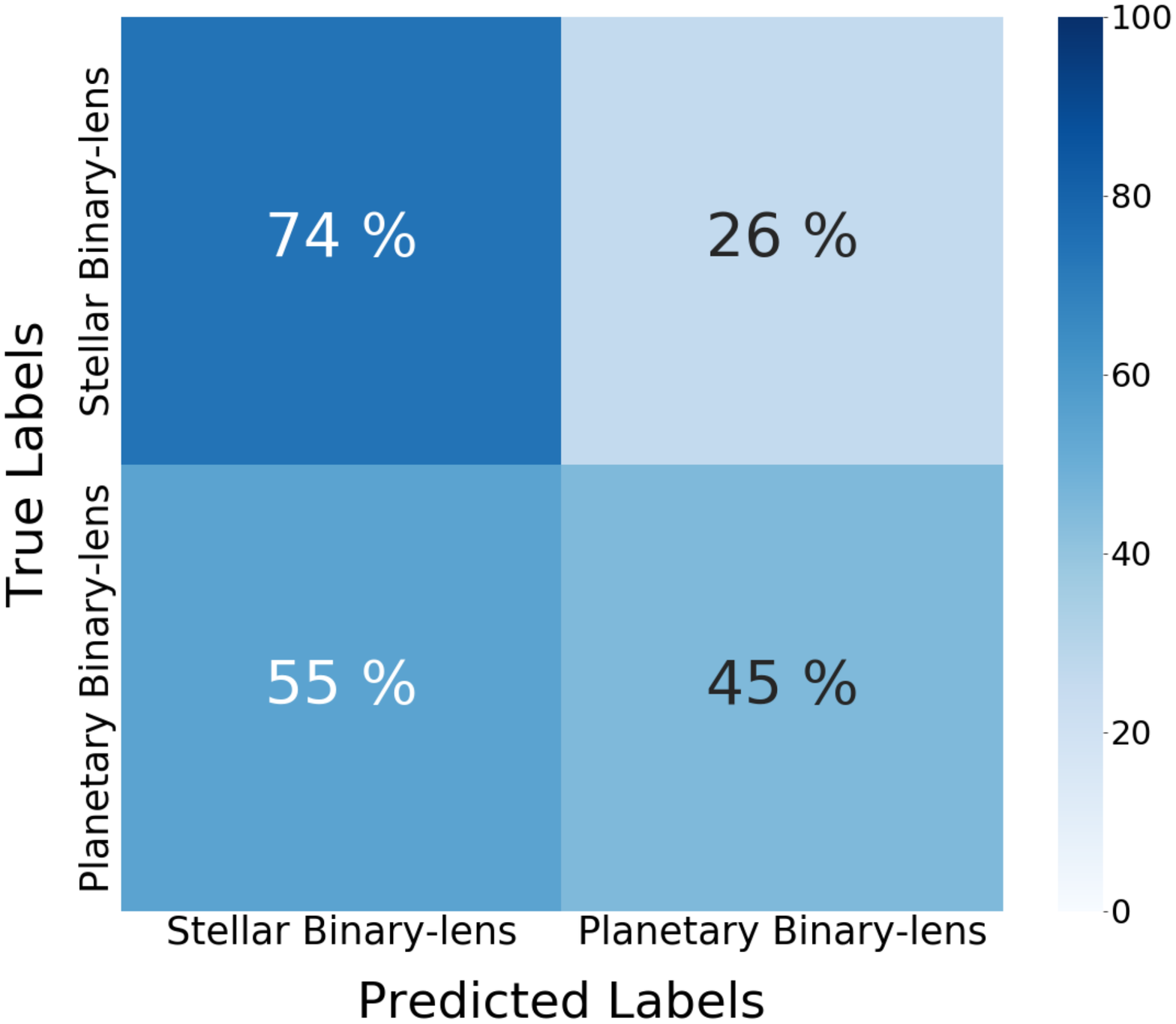}%
      \label{plot:set3_knn}%
    }
    \caption
    {\small Confusion matrices of the four trained classifiers of classification Step III. Once the binary-lens microlensing light curves are found, we use this classification step to classify them into stellar and planetary binary-lens microlensing light curves.} 
    \label{fig:CM_set3}
    
    \end{figure*}

\begin{figure}[t]
    \centerline{\includegraphics[width=1.2\linewidth, clip]{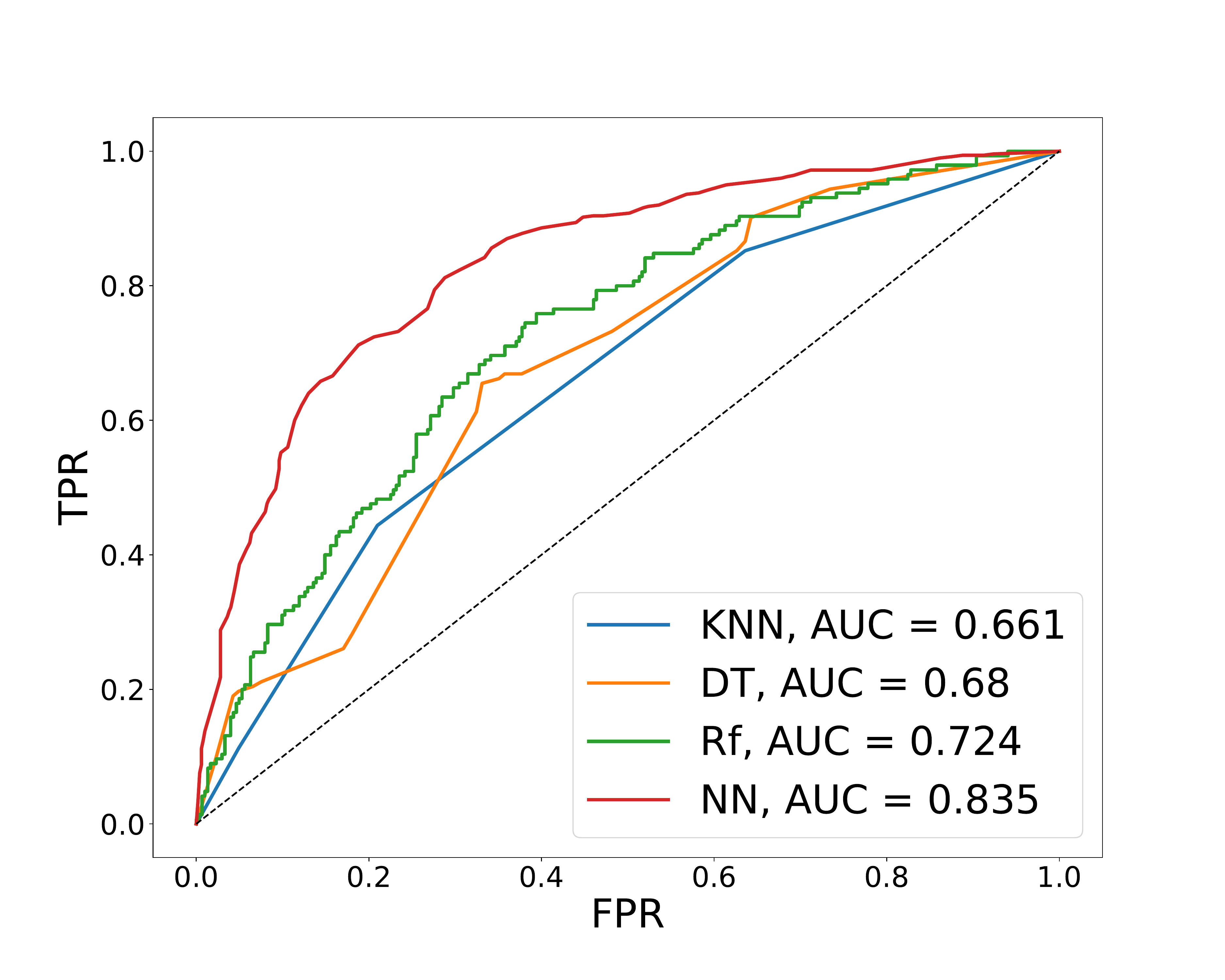}}
    \caption
    {\small ROC curves of the four trained classifiers of classification Step III along with their AUC values. In this step, we aim at classifying binary-lens light curves into two classes of stellar and planetary binary-lens microlensing. Since the dataset is smaller and not well balanced, the overall performance of the classifiers in this step is lower than the other two steps. However, NN shows significantly better performance than the other classifiers. }
    \label{fig:ROC_set3}
    
    \end{figure}

Figure \ref{fig:CM_set3} shows confusion matrices of the four classifiers. The confusion matrix of the NN shows a larger value of TPR compared to all the other confusion matrices, and this is more favorable since our ultimate goal is to detect planetary microlensing light curves. Figure  \ref{fig:ROC_set3} shows the ROC curves and AUC values of the four classifiers in step III. The dataset in this step is smaller compared to the other two steps and is not well balanced. Therefore, the performance of the different classifiers are not as well-differentiated as in the other steps. However, NN shows significantly better performance compared to the other classifiers.

\begin{table*}[!htbp]
    \begin{rotatetable*}
	\caption{Classifiers Hyperparameters and Features.}
	\centering

	\begin{tabular}{|l|l|l|l|l|l|}
	\hline
		{} & {\bf Features} & {\bf KNN} & {\bf DT} & {\bf RF} &{\bf NN}\\ \hline\hline
		 Step I & {\hspace*{-10mm}\begin{tabular}[c]{@{}l@{}} $\Lambda$, $a_2$, $a_4$, $a_6$, $\beta$,\\ Number of peaks \\ (binsize = 60,\\ threshold = 4 \& 6),\\ ${\chi}^2_{PSPL}$ , $t_{E,PSPL}$\end{tabular}} &  {\hspace*{-8mm}\begin{tabular}[c]{l} metric = ``manhattan'',\\ n\_neighbors = 4 \end{tabular}} & {\hspace*{-10mm}\begin{tabular}[c]{@{}l@{}} criterion = ``gini'',\\ max\_depth = 5,\\ splitter = ``best'' \end{tabular}}&  {\hspace*{-10mm}\begin{tabular}[c]{@{}l@{}} max\_features = 7,\\ min\_samples\_leaf = 1\\ 
		n\_estimators = 100,\\ criterion = ``entropy'',\\
		bootstrap = False \end{tabular}} & {\hspace*{-10mm}\begin{tabular}[c]{@{}l@{}}4 Dense Layers,\\ 3 Dropout Layers,\\ activation = ``relu''\\ batch = 32,\\ epochs = 300 \end{tabular}}\\ \hline
		Step II & {\hspace*{-10mm}\begin{tabular}[c]{@{}l@{}} $\Lambda$, $a_2$, $a_4$, $a_6$, $\beta$,\\ ${\chi}^2_{PSPL}$ , $t_{E,PSPL}$\end{tabular}} &  {\hspace*{-8mm}\begin{tabular}[c]{l} metric = ``euclidean'',\\ n\_neighbors = 20 \end{tabular}} & {\hspace*{-10mm}\begin{tabular}[c]{@{}l@{}} criterion = ``entropy'',\\ max\_depth = 5,\\ splitter = ``best'' \end{tabular}}&  {\hspace*{-10mm}\begin{tabular}[c]{@{}l@{}} max\_features = 7,\\ min\_samples\_leaf = 1\\ 
		n\_estimators = 100,\\ max\_depth = 5 \end{tabular}} & {\hspace*{-10mm}\begin{tabular}[c]{@{}l@{}}3 Dense Layers,\\ 2 Dropout Layers,\\ activation = ``relu''\\ batch = 16,\\ epochs = 300 \end{tabular}}\\ \hline
		Step III & {\hspace*{-10mm}\begin{tabular}[c]{@{}l@{}} $\Lambda$, $a_2$, $a_4$, $a_6$,\\ ${\chi}^2_{PSPL}$, $t_{E,PSPL}$, $s$, $q$\end{tabular}} &  {\hspace*{-8mm}\begin{tabular}[c]{l} metric = ``minkowski'',\\ n\_neighbors = 3, p = 15 \end{tabular}} & {\hspace*{-10mm}\begin{tabular}[c]{@{}l@{}} criterion = ``gini'',\\ max\_depth = 7,\\ splitter = ``best'' \end{tabular}}&  {\hspace*{-10mm}\begin{tabular}[c]{@{}l@{}} max\_features = 8,\\ min\_samples\_leaf = 1\\ 
		n\_estimators = 300,\\ max\_depth = 8 \end{tabular}} & {\hspace*{-10mm}\begin{tabular}[c]{@{}l@{}}4 Dense Layers,\\ 2 Dropout Layers,\\ activation = ``tanh''\\ batch = 32,\\ epochs = 100 \end{tabular}}\\ \hline
		 \end{tabular}
	\label{table:classifiers_info}
	\end{rotatetable*}
\end{table*}

\section{Conclusion and Future Work}\label{sec:conclusion}

Classifying light curves that manifest different types of stellar variability is still a major challenge. Although using ML methods to classify astronomical time series is a powerful tool, it is important to understand which method we should choose and what are the steps we need to take in order to use these methods effectively. 

In this paper, we introduce a new approach towards classifying microlensing light curves based on light curve morphologies. We introduce a package of tools including several functional fits that can be applied to the light curve in a fast and efficient way to extract information about the different morphological features in the light curves. This information is quantified as features that can be used to make decisions about the light curve types. 

This approach will be useful when it comes to analyzing large microlensing data sets from high-cadence surveys like Roman \citep{spergel2015wide} in the future and ongoing surveys like KMTNet \citep{kim2010technical}. Our preliminary results in Section \ref{sec:ML_test} show that the features produced by our tools can be used as input to ML classifiers like Random Forest to distinguish between different types of microlensing light curves, and help prioritize the ones that are more likely to be caused by planetary systems. 

The simulated data used in this paper included CV-like events as the only non-microlensing instance. An ideal dataset would include a variety of stellar variability, and our developed package should be modified to include all other variability in its analysis. We believe that the same approach will be successful in recovering microlensing events from a wide variety of non-microlensing variability on its own. 

There are currently a number of methods that attempt to categorize photometric variability in large data sets, such as parametric statistical methods and ML. ML methods for detecting different types of variabilities are becoming more common \citep[e.g.][]{richards2011machine,pichara2013automatic,pashchenko2017machine,valenzuela2017unsupervised}. For example, as mentioned before, \citet{godines2019machine} has trained a Random Forest classifier to detect microlensing light curves in a low-cadence survey dataset in real time. Our produced features can be a complementary set of features for such algorithms and not only can improve their results but also can give them the ability to detect possible planetary binary-lenses as well.

In ML classifiers, increasing the number of object instances can significantly improve the results. Our largest training set included 4181 light curves, which is not a large number for most ML applications. Increasing this number to about $\sim 10,000$ light curves would yield more robust and reliable results. Some of the tools presented in this paper produce other parameters as well, and including different subsets of those parameters could also improve the results. This task needs to be done carefully, though, since adding more features that are not important might result in overfitting. Additionally, a great avenue to improve the results would be to test other classifiers like the Support Vector Classifier and Naive Bayesian, and also investigate deeper neural networks. The improvement of the data set and the ML algorithms will be presented in a second paper of this series.

\section{Acknowledgement}

S.K thanks the LSSTC Data Science Fellowship Program, which is funded by LSSTC, NSF Cybertraining Grant $\#1829740$, the Brinson Foundation, and the Moore Foundation; her participation in the program has benefited this work. She also thanks the International Microlensing Conference that took place in 2018 in New York City, NY, where there were discussions and works that initiated this project. R.S gratefully acknowledges support from NASA grant 80NSSC19K0291. B.S.G. was supported by a Thomas Jefferson Chair for Space Exploration endowment at the Ohio State University.

\bibliographystyle{apj}
\bibliography{References.bib}
\clearpage
 
\end{document}